\documentclass[%
 reprint,
superscriptaddress,
 amsmath,amssymb,
 aps,
]{revtex4-2}

\usepackage{graphicx}% Include figure files
\usepackage{dcolumn}% Align table columns on decimal point
\usepackage{bm}% bold math
\usepackage{siunitx}[=v3]
\usepackage{import}
\begin{document}

%\preprint{APS/123-QED}

\title{High Coherence in a Tileable 3D Integrated Superconducting Circuit Architecture}% Force line breaks 
\author{Peter A Spring}
% \altaffiliation[Also at ]{Physics Department, XYZ University.}%Lines break automatically or can be forced with \\
\author{Shuxiang Cao}
\author{Takahiro Tsunoda}
\author{Giulio Campanaro}
\author{Simone D Fasciati}
\author{James Wills}
\author{Vivek Chidambaram}
\author{Boris Shteynas}
\author{Mustafa Bakr}
\affiliation{%
Department of Physics, Clarendon Laboratory, University of Oxford, Oxford, OX1 3PU, United Kingdom
}
\author{Paul Gow}
\author{Lewis Carpenter} 
\author{James Gates}
\affiliation{
Optoelectronics Research Centre, University of Southampton, Southampton, SO17 1BJ, 
United Kingdom
}%
\author{Brian Vlastakis}
\author{Peter J Leek}%
\email{peter.leek@physics.ox.ac.uk}
\affiliation{%
Department of Physics, Clarendon Laboratory, University of Oxford, Oxford, OX1 3PU, United Kingdom
}
 \email{peter.leek@physics.ox.ac.uk}

\date{\today}% It is always \today, today,
             %  but any date may be explicitly specified
\begin{abstract}
We report high qubit coherence as well as low crosstalk and single-qubit gate errors in a superconducting circuit architecture that promises to be tileable to 2D lattices of qubits. The architecture integrates an inductively shunted cavity enclosure into a design featuring non-galvanic out-of-plane control wiring and qubits and resonators fabricated on opposing sides of a substrate. The proof-of-principle device features four uncoupled transmon qubits and exhibits average energy relaxation times $T_1=\SI{149(38)}{\micro \second}$, pure echoed dephasing times $T_{\phi,e}=\SI{189(34)}{\micro \second}$, and single-qubit gate fidelities $F=99.982(4)\%$ as measured by simultaneous randomized benchmarking. The 3D integrated nature of the control wiring means that qubits will remain addressable as the architecture is tiled to form larger qubit lattices. Band structure simulations are used to predict that the tiled enclosure will still provide a clean electromagnetic environment to enclosed qubits at arbitrary scale.
\end{abstract}
\maketitle
\section{\label{sec:level1} Introduction}
Building 2D lattices of hundreds or thousands of individually addressable, highly coherent qubits is an outstanding hardware challenge. Anticipated applications include demonstrations of logical gates using the surface code~\cite{Andersen2020,Barends2014, Fowler2012, Bravyi1998} and quantum simulations of 2D lattice Hamiltonians~\cite{Yanay2020,Gong2021}. Superconducting circuits are a promising platform for realizing such lattices~\cite{Gong2021, Arute2019, Otterbach2017}; qubits are lithographically defined on 2D substrates, and tailored coupling circuitry can be included in the regions between qubits to realize a universal gate set. Two requirements for scaling such superconducting qubit lattices are: (1) a method to route control wiring to the circuit such that all qubits remain addressable and measurable at progressively larger scales; and (2) a means of preventing low frequency spurious modes from emerging in the circuit as the dimensions increase~\cite{Kjaergaard2020}. These modes can arise from sections of spurious planar transmission lines such as slotlines~\cite{Chen2014,Huang2021}, or from 3D cavity enclosures that house the circuit~\cite{Huang2021}. Solutions to these scaling challenges must not introduce significant decoherence channels to qubits; and if fault tolerance is desired, must be compatible with gate fidelities beyond the thresholds of quantum error correction codes.
\\
To overcome the wiring limitations of edge-connected circuits, 3D integrated control wiring is a practical solution. Various approaches to this have been demonstrated, for example with spring-loaded pogo pins \cite{Bejanin2016,Bronn2018} and with galvanic bonding of the qubit substrate to a wiring/interposer substrate~\cite{Foxen2017,Yost2020,Rosenberg2017}. To avoid spurious modes due to slotlines, divided ground planes can be inductively shunted with airbridges~\cite{Chen2014} or with superconducting through substrate vias (TSVs)~\cite{AlfaroBarrantes2020, Yost2020,Vahidpour2017}. To avoid low frequency cavity modes, one solution is to divide the quantum processor into subsystems, with each subsystem enclosed in a cavity with dimensions $\lesssim
\SI{1}{\centi\meter}$~\cite{Lei2020, Brecht2017, Brecht2016, Brecht2015}. Alternatively, circuits can be enclosed in inductively shunted cavities that can scale arbitrarily in two dimensions with a cutoff frequency to cavity modes~\cite{Spring2020, Murray2016}.
\\
In this work, we present experimental results on a four-qubit proof-of-principle circuit incorporating this latter concept. The circuit architecture, based on that introduced in Refs.~\cite{Rahamim2017, Patterson2019}, features 3D integrated out-of-plane control wiring and qubits and readout resonators that are fabricated on opposing sides of a substrate. We incorporate a key new feature: inductively shunting the circuit enclosure with a CNC machined pillar that passes through the substrate. This design is established to be compatible with transmon coherence times exceeding \SI{100}{\micro\second}, as well as low crosstalk and single-qubit gate errors. Simulations of band structure are used to predict that 2D qubit lattices can be formed by tiling a unit cell within the architecture, without the emergence of low frequency cavity modes and with exponentially decaying cavity mediated crosstalk between qubits.
%\vspace{-0.1em}
\section{Device Architecture}
\begin{figure*}
\includegraphics[width=\textwidth,height=\textheight,keepaspectratio]{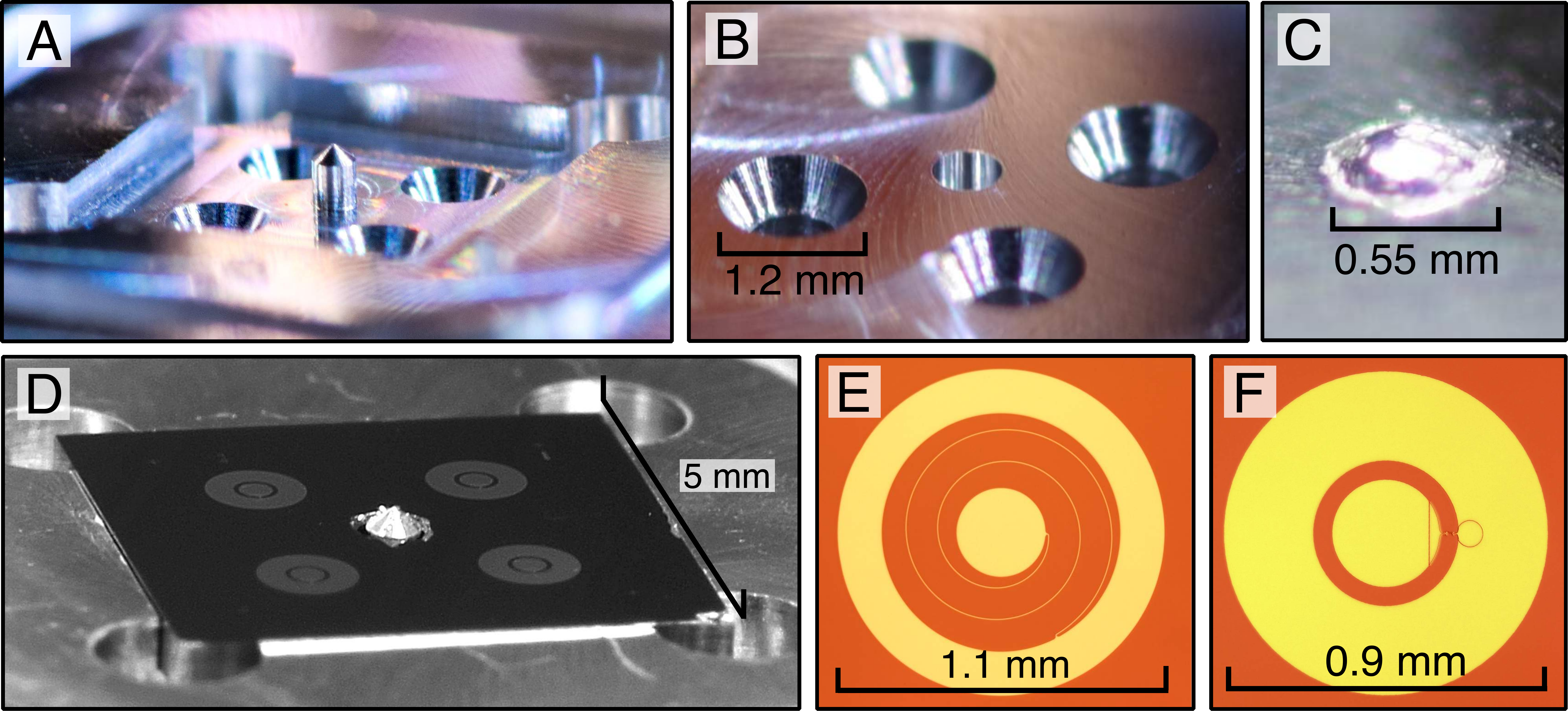}% Here is how to import EPS art
\caption{\label{fig:Device_optical_images} \textbf{Optical images of cavity enclosure and circuit.} (\textbf{A}) Enclosure base with cavity, central pillar, and four tapered through-holes for out-of-plane wiring access. (\textbf{B}) Enclosure lid with a central cylindrical recess and identical through-holes for out-of-plane wiring. (\textbf{C}) Cylindrical recess in the lid filled with a ball of indium. (\textbf{D}) (Grayscale) Four-qubit circuit mounted inside the enclosure base. The four qubits are visible, arranged in a square lattice with $\SI{2}{\milli\meter}$ spacing. (\textbf{E})-(\textbf{F}) A spiral resonator and a transmon qubit with identical electrode dimensions to those in the device. }
\end{figure*}

\begin{figure*}
\includegraphics[width=0.9\textwidth,height=0.9\textheight,keepaspectratio]{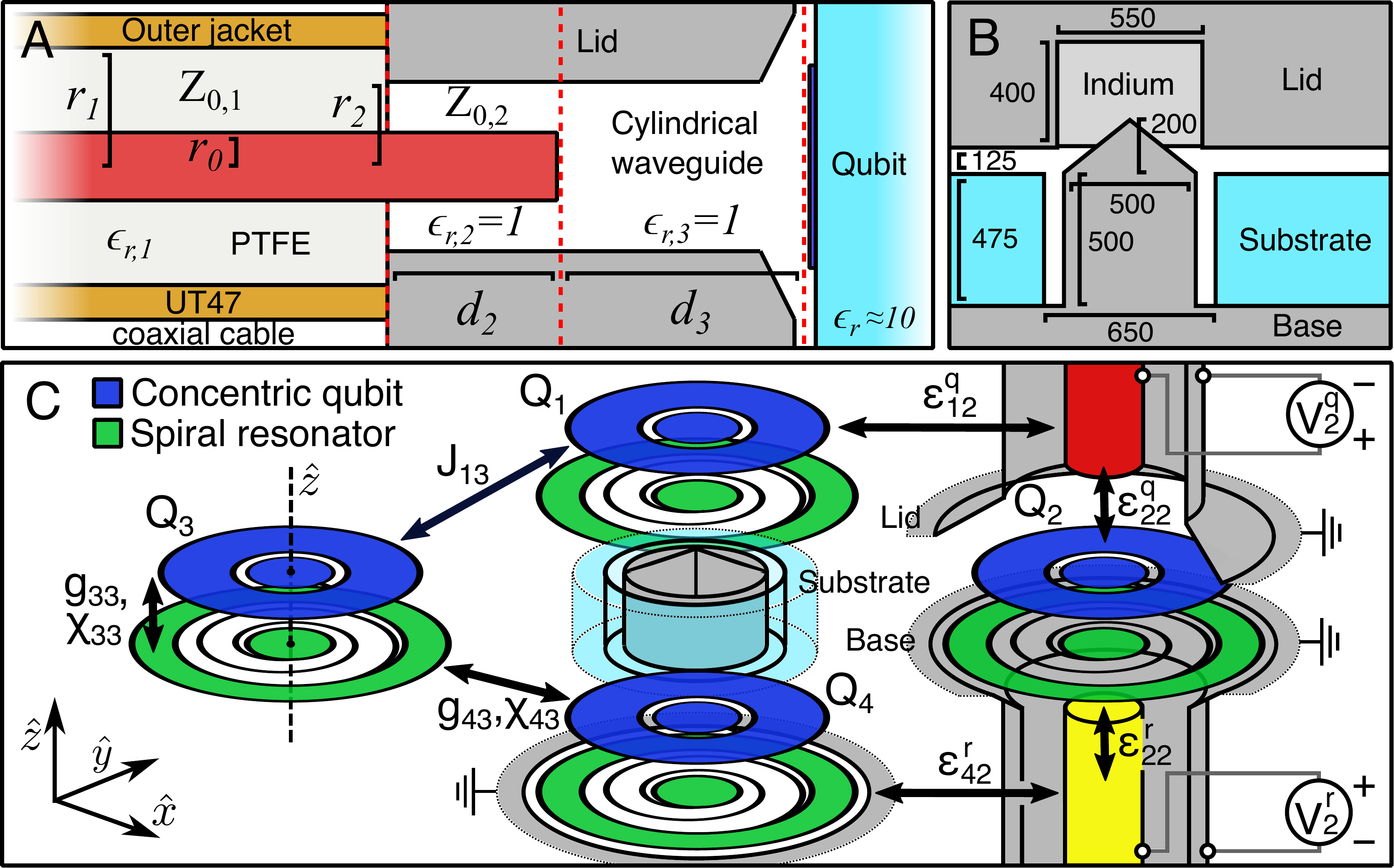}
\caption{\label{fig:Device_schematics} \textbf{Device schematics.} (\textbf{A}) Cross-section of the out-of-plane wiring design (not to scale), here shown addressing a qubit. (\textbf{B}) Cross-section of the ‘bulk via’ inductive shunt design (to scale). The designed dimensions are shown in µm. (\textbf{C}) Circuit layout illustration (not to scale). The substrate and cavity enclosure are partially shown, and the out-of-plane wiring is shown for $Q_2$. Examples of the coupling terms and drive voltages in the Hamiltonian in eqs.~\ref{eq:basic_Ham}-\ref{eq:crosstalk_Ham} are shown.}
%\vspace{-0.3em}
\end{figure*}

Fig.~\ref{fig:Device_optical_images} shows optical images of the cavity enclosure and circuit. 
The enclosure base [Fig.~\ref{fig:Device_optical_images}(A)] features a single central ‘pillar’, and the lid [Fig.~\ref{fig:Device_optical_images}(B)] contains a matching cylindrical recess that is filled with a ball of indium [Fig.~\ref{fig:Device_optical_images}(C)]. The base and lid both contain four tapered through-holes that act as waveguides for qubit and resonator control signals. In Fig.~\ref{fig:Device_optical_images}(D), the circuit substrate is shown placed inside the enclosure base. An aperture has been machined in the center of the substrate allowing the pillar to pass through. The four coaxial transmon~\cite{Koch2007} qubits are visible, arranged in a $2\times2$ lattice with \SI{2}{\milli\meter} spacing. 
\\
Fig.~\ref{fig:Device_schematics} shows a schematic of the out-of-plane wiring design, the inductive shunt design, and the circuit layout. Control signals are routed to qubits and resonators by UT47-type coaxial cables with characteristic impedance $Z_{0,1}=50\pm \SI{2.5}{\ohm}$. As shown in Fig.~\ref{fig:Device_schematics}(A), the inner conductor of each cable, radius $r_0$, extends a distance $d_2$ into a through-hole in the circuit base/lid part, radius $r_2$, forming a coaxial waveguide with characteristic impedance $Z_{0,2}$. In this device, $r_0=145\pm\SI{2.5}{\micro \meter}$ and $r_2=350\pm\SI{10}{\micro\meter}$ (at room temperature) such that when ideally aligned $Z_{0,2}=52.8\pm\SI{2.5}{\ohm}$, a close match to the coaxial cable impedance $Z_{0,1}$. The PTFE in the coaxial cable is separated from the circuit substrate by approximately \SI{5}{\milli\meter}, reducing qubit/resonator electric field participation~\cite{wang2015surface} in this lossy dielectric. The inner conductor of the coaxial cable terminates a distance $d_3$ from the qubit/resonator that it addresses. The coupling of control signals to qubits/resonators is dominantly mediated by an evanescent $\text{TM}_{01}$ circular waveguide mode~\cite{Reagor2016}, with coupling strength $\varepsilon \propto e^{-d_3/\delta_{c}}$, $\delta_{c}\approx r_2/2.4$. For $r_2=\SI{350}{\micro \meter}$, $\delta_{c}\approx\SI{150}{\micro \meter}$. In this device, $d_3 \approx \SI{0.9}{\milli\meter}$ $(\SI{0.4}{\milli \meter})$ for each qubit (resonator) control line. Fig.~\ref{fig:Device_schematics}(B) shows a schematic cross-section of the pillar in the enclosure base, passing through the aperture in the circuit substrate and galvanically connecting to the indium filled recess in the enclosure lid. The pillar acts as a ‘bulk via’ that inductively shunts the two halves of the enclosure, without requiring side wall metallization of the substrate aperture or a galvanic connection between the substrate and enclosure. Fig.~\ref{fig:Device_schematics}(C) shows the circuit layout. The reverse side of the substrate rests directly on the enclosure base and contains four lumped LC ‘spiral’ resonators. Each resonator is coaxially aligned with and capacitively coupled to a qubit. The cavity enclosure provides the ground, and there are no ground planes on the substrate. 
The qubit (resonator) electrodes are electrically floating and are designed to have a shortest distance to the surface of the cavity enclosure of approximately \SI{200}{\micro \meter} (\SI{50}{\micro \meter}).
\begin{table*}
\caption{\label{tab:basic_characterization} \textbf{Basic device characterization.} Summary of basic circuit parameters. The quantities $\omega_q$, $\omega_r$, $\alpha$, $\chi$ and $g$ are as defined in the main text. The quantity $E_{J,i}/E_{C,i}$ is the ratio of the Josephson energy to the charging energy in qubit $i$, $\kappa_{ext,i}$ is the external decay rate of resonator $i$, $Q_{int,i}$ is the internal quality factor of resonator $i$, and $p_{e,i}$ is the residual excited state population in qubit $i$.}
\begin{ruledtabular}
\begin{tabular}{cccccccccc}
&$\omega_q/2\pi$ $(\si{\giga\hertz})$ & $\omega_r/2\pi$ $(\si{\giga\hertz})$ & $\alpha/2\pi$ $(\si{\mega\hertz})$
& $E_J/E_C$ & $\chi/2\pi$ $(\si{\kilo\hertz})$ & $g/2\pi$ $(\si{\mega\hertz})$ & $\kappa_{ext}/2\pi$ $(\si{\kilo\hertz})$ & $Q_{int}$ $(10^3)$ & $p_e$ $(\%)$\\ \hline
 $Q_1/R_1$&3.981&7.968&-199&69&-165&124&118&110&13 \\
 $Q_2/R_2$&4.045&8.083&-199&71&-167&126&73&75&18\\
 $Q_3/R_3$&4.130&8.183&-198&74&-169&128&749&515&13\\
 $Q_4/R_4$&4.192&8.289&-197&76&-164&128&241&160&10\\
\end{tabular}
\end{ruledtabular}
\end{table*}
\section{Basic characterization}
\begin{figure*}
\includegraphics[width=0.85\textwidth,height=0.85\textheight,keepaspectratio]{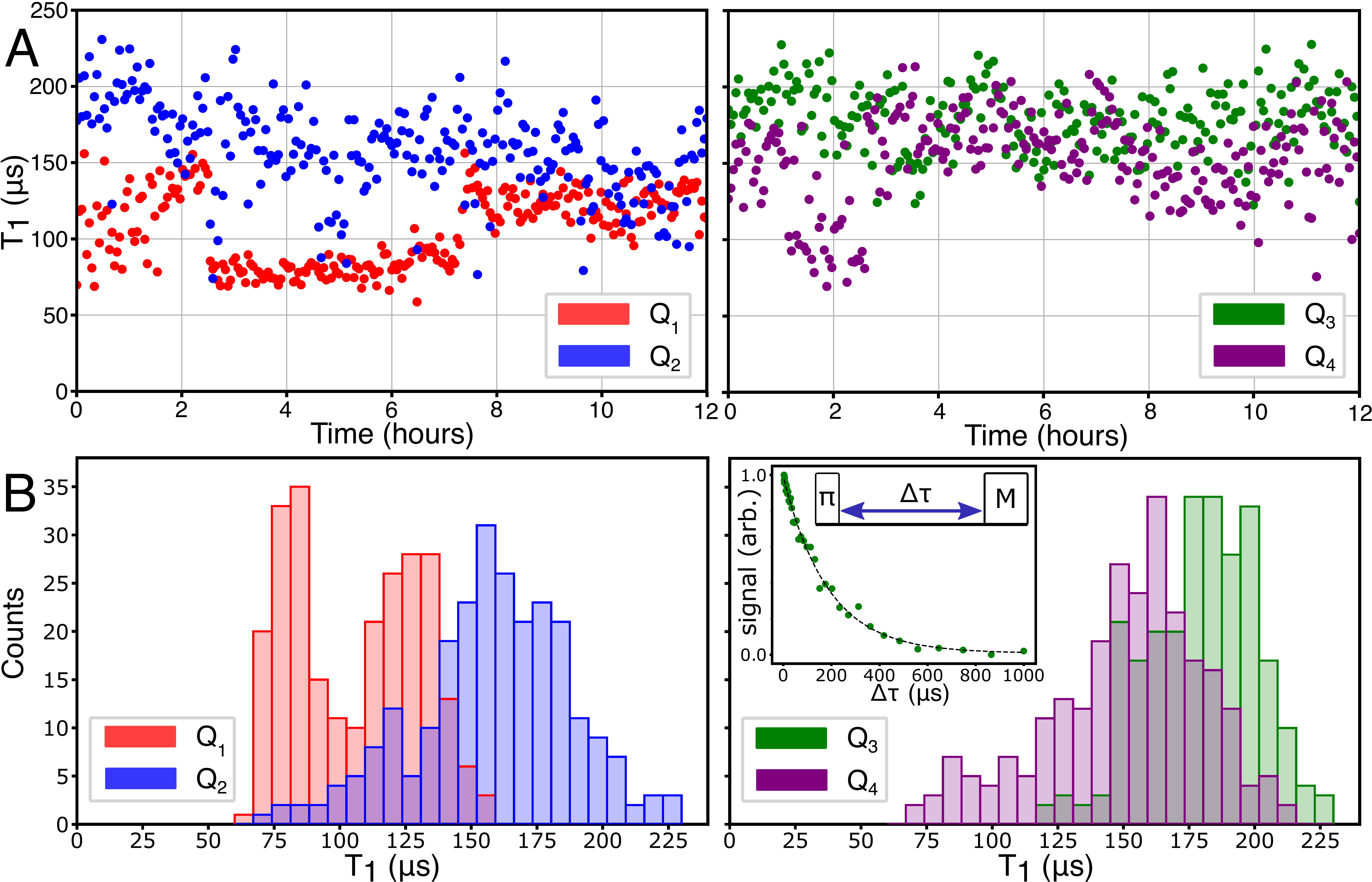}
\caption{\label{fig:Qubit_relaxation_times} \textbf{Qubit relaxation characterization.} (\textbf{A}) 251 consecutive $T_1$ measurements over an approximately 12-hour period. (\textbf{B}) Resultant histograms of $T_1$. The inset shows an example $T_1$ time trace for $Q_3$, and the measurement pulse sequence. The four qubits were measured simultaneously; the data is shown across two graphs for legibility.}
\end{figure*}

An effective dispersive Hamiltonian for the low energy spectrum of the device is given by
%\begin{widetext}
\begin{eqnarray}
\hat{H}/\hbar =  \sum_{i=1}^{4} \omega_{q,i}\hat{a}^\dag_i\hat{a}_i +\frac{\alpha_i}{2}\hat{a}^\dag_i\hat{a}_i(\hat{a}^\dag_i\hat{a}_i-1) + \omega_{r,i}\hat{b}^\dag_i\hat{b}_i+ \nonumber\\
2\chi_{ii}\hat{a}^\dag_i\hat{a}_i\hat{b}^\dag_i\hat{b}_i + \varepsilon_{ii}^q(\hat{a}_i-\hat{a}^\dag_i)V_i^q + \varepsilon_{ii}^r(\hat{b}_i-\hat{b}^\dag_i)V_i^r + \frac{\hat{H}_s}{\hbar}\text{.} \; \; \; \;
\label{eq:basic_Ham}
\end{eqnarray}
%\end{widetext}
Here, $\hat{a}^\dag_i$ ($\hat{a}_i$) and $\hat{b}^\dag_i$ ($\hat{b}_i$) are the creation (annihilation) operators for qubit $i$ and resonator $i$ respectively; $\omega_{q,i}$ is the transition frequency of qubit $i$ given zero photons in resonator $i$; $\alpha_i$ is the anharmonicity of qubit $i$; $\omega_{r,i}$ is the frequency of resonator $i$ given qubit $i$ is in its ground state; $\chi_{ii}$ is the dispersive shift between qubit $i$ and resonator $i$; and $\varepsilon_{ii}^q$ ($\varepsilon_{ii}^r$) describe the coupling of qubit (resonator) $i$ to qubit (resonator) control line $i$, which is driven with a voltage $V_i^q$ ($V_i^r$). These voltages are applied close to the cylindrical waveguide transition and at a fixed distance from the circuit [see Fig.~\ref{fig:Device_schematics}(C)]. $\hat{H}_s$ contains undesired crosstalk terms that are discussed in the crosstalk characterization section.
\\
The quantities $\omega_{q,i}$, $\alpha_i$, $\omega_{r,i}$ and $\chi_{ii}$ were determined using standard spectroscopic measurements and Ramsey measurements, and are shown in Table~\ref{tab:basic_characterization}. The relaxation times $T_1$ of the four qubits were simultaneously measured repeatedly over a period of 12 hours. The consecutive measured values and resulting histograms are shown in Fig.~\ref{fig:Qubit_relaxation_times}. The characteristic dephasing times $T_2^*$ and $T_{2,e}$ were measured using standard Ramsey and Hahn echo pulse sequences (see Appendices). The coherence times are summarized in Table~\ref{tab:coherence_characterization}.
\begin{table*}
\caption{\label{tab:coherence_characterization} \textbf{Qubit coherence and gate errors.} Summary of coherence results and error-per-physical gate (EPG) as found by separate (sep) and simultaneous (sim) randomized benchmarking. The pure echoed dephasing times are given by $1/T_{\phi,e}=1/T_{2,e} - 1/(2T_1)$. The coherence limited EPG (EPG coh lim) was calculated as $(3-\text{exp}(-\tau_g/T_1)-2\text{exp}(-\tau_g/T_{2,e}))/6$~\cite{IBM}, where $\tau_g$ is the total period of each physical gate, here $\SI{24}{\nano\second}$.}
\begin{ruledtabular}
\begin{tabular}{cccccccc}
& $T_1$ $(\si{\micro\second})$&$T_2^*$ $(\si{\micro\second})$ & $T_{2,e}$ $(\si{\micro\second})$ & $T_{\phi,e}$ $(\si{\micro\second})$ & EPG sep $(10^{-4})$ & EPG sim $(10^{-4})$ & EPG coh lim $(10^{-4})$ \\ \hline
 $Q_1$&106(24)&95(5)&101(9)&193(52)&2.29(4)&1.64(4)&1.1(1)\\
 $Q_2$&159(30)&104(9)&116(6)&183(25)&1.46(6)&2.15(8)&0.94(5)\\
 $Q_3$&179(21)&89(12)&128(9)&199(25)&1.16(5)&1.31(5)&0.85(5)\\
 $Q_4$&151(30)&99(8)&113(4)&181(24)&2.23(4)&2.16(4)&0.97(5)\\
 Avg.&149(38)&97(10)&115(12)&189(34)&1.8(5)&1.8(4)&1.0(1)\\
\end{tabular}
\end{ruledtabular}
\end{table*}
\begin{figure*}
\includegraphics[width=0.85\textwidth,height=0.85\textheight,keepaspectratio]{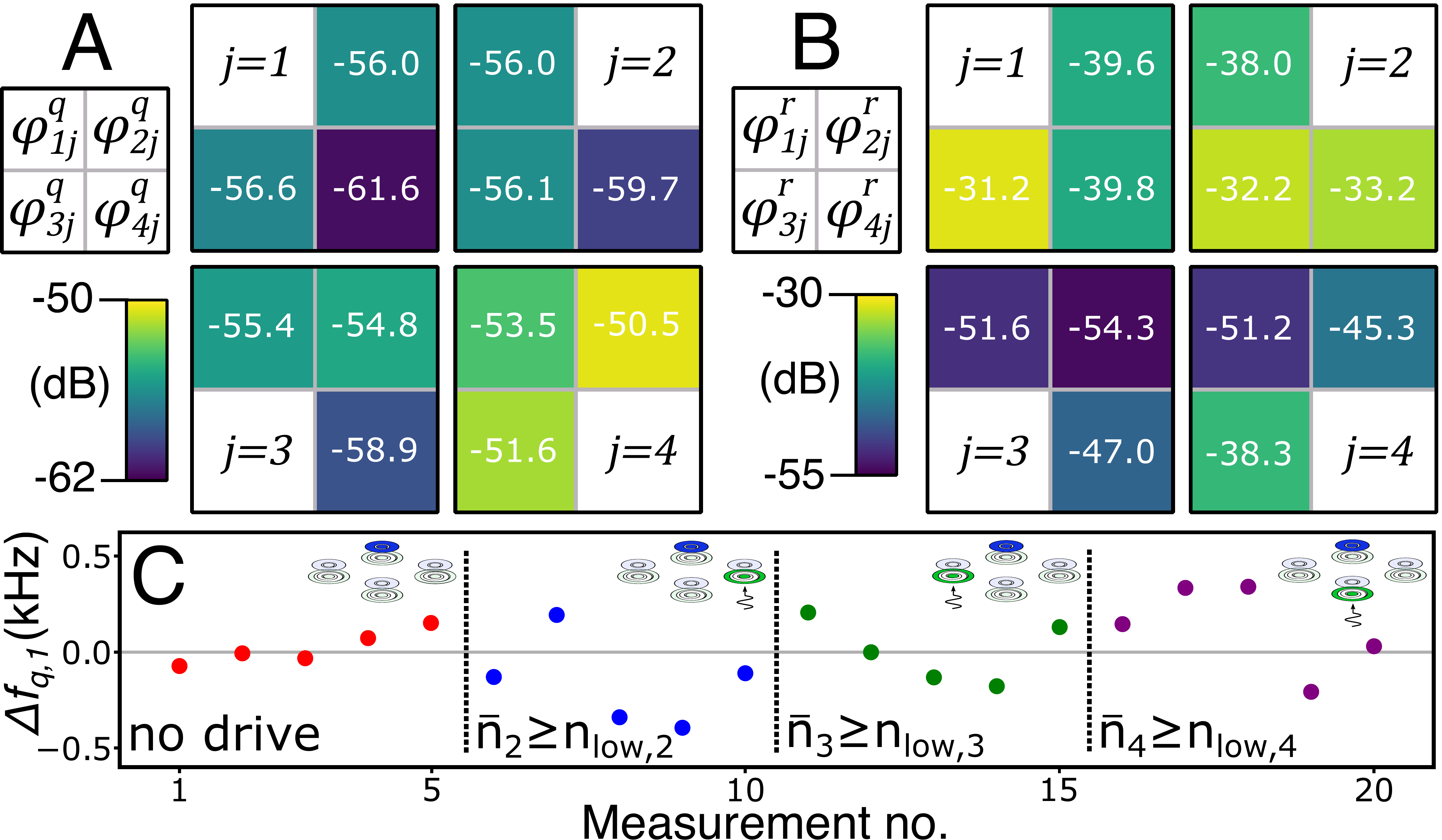}
\caption{\label{fig:crosstalk_characterization} \textbf{Crosstalk characterization.} (\textbf{A}) Experimentally measured qubit control line selectivity $\varphi_{ij}^q=(\varepsilon_{ij}^q/\varepsilon_{jj}^q)^2$ from qubit $i$ to qubit control line $j$, expressed in units of $\si{\decibel}$ as $10\log_{10}(\varphi^q_{ij})$. (\textbf{B}) Experimentally measured resonator control line selectivity $\varphi_{ij}^r=(\varepsilon_{ij}^r/\varepsilon_{jj}^r)^2$ from resonator $i$ to resonator control line $j$, expressed in units of $\si{\decibel}$ as $10 \log_{10}(\varphi^r_{ij})$. (\textbf{C}) Frequency variation in $Q_1$ found from 20 repeated Ramsey experiments; with either no drive on any resonator, or a continuous drive applied to $R_2$, $R_3$, or $R_4$ at frequency $\omega_{r,j}$ that populates it with a photon number $\bar{n}_j$ of at least $n_{low,j} \stackrel{\text{def}}{=} n_{crit,j}/10$.}
\end{figure*}
\section{Crosstalk characterization}
The device is a proof-of-principle demonstration of the circuit architecture with no intentional couplings except between qubit-resonator pairs; as such, we identify all other couplings as undesired crosstalk. The crosstalk terms that were considered are defined in the following effective Hamiltonian:
\begin{eqnarray}
\hat{H}_s/\hbar =  \sum_{i,j(i\neq j)}^{4} J_{ij}\hat{a}^\dag_i\hat{a}_j+2\chi_{ij}\hat{a}^\dag_i\hat{a}_i\hat{b}^\dag_j\hat{b}_j+ \; \; \; \; \; \; \; \; \; \; \nonumber\\
\varepsilon_{ij}^q(\hat{a}_i-\hat{a}^\dag_i)V_j^q + \varepsilon_{ij}^r(\hat{b}_i-\hat{b}^\dag_i)V_j^r\text{.}
\label{eq:crosstalk_Ham}
\end{eqnarray}
Here, $J_{ij}$ is a parasitic transverse coupling between qubits $i$ and $j$, satisfying $J_{ij}=J_{ji}$; $\chi_{ij}$ is a parasitic dispersive shift between qubit $i$ and resonator $j$; and $\varepsilon_{ij}^q$ ($\varepsilon_{ij}^r$) describes a parasitic coupling between qubit (resonator) $i$ and qubit (resonator) control line $j$. Some examples of these different types of crosstalk are shown pictorially in Fig.~\ref{fig:Device_schematics}(C). The following expression was used to relate the dispersive shift $\chi_{ij}$ to a transverse coupling $g_{ij}$ between transmon qubit $i$ and resonator $j$~\cite{Koch2007}
\begin{equation}
\chi_{ij}\approx-\frac{(g_{ij})^2 E_{C,i}/\hbar}{\Delta_{ij} (\Delta_{ij}-E_{C,i}/\hbar)}\text{,}
\end{equation}
where $\Delta_{ij}=\omega_{q,i}-\omega_{r,j}$ and $E_{C,i}$ is the charging energy of qubit $i$.
\\
The experimentally bounded maximum parasitic transverse couplings are summarized in Table~\ref{tab:crosstalk_bounds}, along with the predicted maximum values found by applying a simple impedance formula~\cite{Solgun2019} to HFSS~\cite{AnsysHFSS} driven terminal simulations (see Appendices).
\subsection{Qubit control line selectivity}
\begin{table}[b]
\caption{\label{tab:crosstalk_bounds} \textbf{Bounds on parasitic transverse couplings.} Experimentally determined bounds on the magnitude of parasitic transverse couplings in the device, and the maximum predicted values between any qubit-qubit/qubit-resonator pair found using HFSS driven terminal simulations and an impedance formula~\cite{Solgun2019}.}
\begin{ruledtabular}
\begin{tabular}{ccc}
Crosstalk quantity & Experiment & Simulation \\
& $(\si{\kilo\hertz})$ & $(\si{\kilo\hertz})$\\
\hline
Qubit-qubit coupling $|J|/2\pi$ & $<250$ & 10 \\
Qubit-resonator coupling $|g|/2\pi$ & $<1500$ & 50\\
\end{tabular}
\end{ruledtabular}
\end{table}
The qubit control line selectivity $\varphi_{ij}^q$ is here defined
\begin{equation}
\varphi_{ij}^q \stackrel{\text{def}}{=} \left(\frac{\varepsilon_{ij}^q}{\varepsilon_{jj}^q}\right)^2\text{.}
    \label{eq:qubit_selectivity}
\end{equation}
The selectivity was measured by driving qubit $i$ at frequency $\omega_{q,i}$ over a range of generator drive voltages $V_j^{q,gen}$ and fitting the induced Rabi oscillation rate $\Omega_i$ to the linear function $\Omega_i=k_{ij} V_j^{q,gen}$. From the measured linear response in the strong drive regime $|\varepsilon_{jj}^q V_j^q|\gg |\Delta_{ij}^q|$, it is inferred that $|J_{ij}/\Delta_{ij}^q|\ll|\varepsilon_{ij}^q/\varepsilon_{jj}^q|$, where $\Delta_{ij}^{q}=\omega_{q,i}-\omega_{q,j}$ (see Appendices). In this case, the selectivity takes the simple form $\varphi_{ij}^q=(k_{ij}/k_{jj})^2$. The measured qubit control line selectivities are shown in Fig.~\ref{fig:crosstalk_characterization}(A), and the plots of $\Omega_i$ vs. $V_j^{q,gen}$ that were used to determine $\varphi_{21}^q$ are shown in Fig.~\ref{fig:qubit_control_line_selectivity_measurement}. Making use of the fact $|J_{ij}/\Delta_{ij}^q|\ll|\varepsilon_{ij}^q/\varepsilon_{jj}^q|$ in this device results in the following experimental bound on the transverse coupling $J_{ij}$: $|J_{ij}|<[\text{min}(\varphi_{ij}^q, \varphi_{ji}^q)]^{0.5}|\Delta_{ij}^q|$.
\subsection{Resonator control line selectivity}
The resonator control line selectivity $\varphi_{ij}^r$ is here defined 
\begin{equation}
\varphi_{ij}^r \stackrel{\text{def}}{=} \left(\frac{\varepsilon_{ij}^r}{\varepsilon_{jj}^r}\right)^2 \text{.}
    \label{eq:resonator_selectivity}
\end{equation}
The selectivity was measured by continuously driving resonator $i$ at frequency $\omega_{d,i}=\omega_{r,i}+\Delta_d$, where $|\Delta_d|\gg \kappa_i,|\chi_{ii}|$ ($\kappa_i$ is the total decay rate of resonator $i$), over a range of generator drive powers $P_j^{r,gen}$. The induced AC Stark-shift $\omega_{AC,i}$ in qubit $i$ was then fit to the linear function $\omega_{AC,i}=k_{ij}' P_j^{r,gen}$. The selectivity is then given by $\varphi_{ij}^r=(\chi_{jj}/\chi_{ii})(k_{ij}'/k_{jj}')$ (see Appendices). The measured resonator control line selectivities are shown in Fig.~\ref{fig:crosstalk_characterization}(B), and the plots of $\omega_{AC,i}$ vs. $P_j^{r,gen}$ that were used to determine $\varphi_{21}^r$ are shown in Fig.~\ref{fig:resonator_control_line_selectivity_measurement}.
\subsection{Parasitic qubit-resonator coupling}
To measure the parasitic dispersive shift $\chi_{ij}$ between qubit $i$ and resonator $j$, resonator $j$ was continuously driven at frequency $\omega_{r,j}$ from its own control line to populate it with a steady-state photon number $\bar{n}_j$ of at least $n_{crit,j}/10$, where $n_{crit,j}=(\Delta_{jj}/2g_{jj})^2$ is the critical photon number~\cite{Blais2020Circuit} of resonator $j$ (see Appendices).
\begin{figure*}
\includegraphics[width=0.98\textwidth,height=0.98\textheight,keepaspectratio]{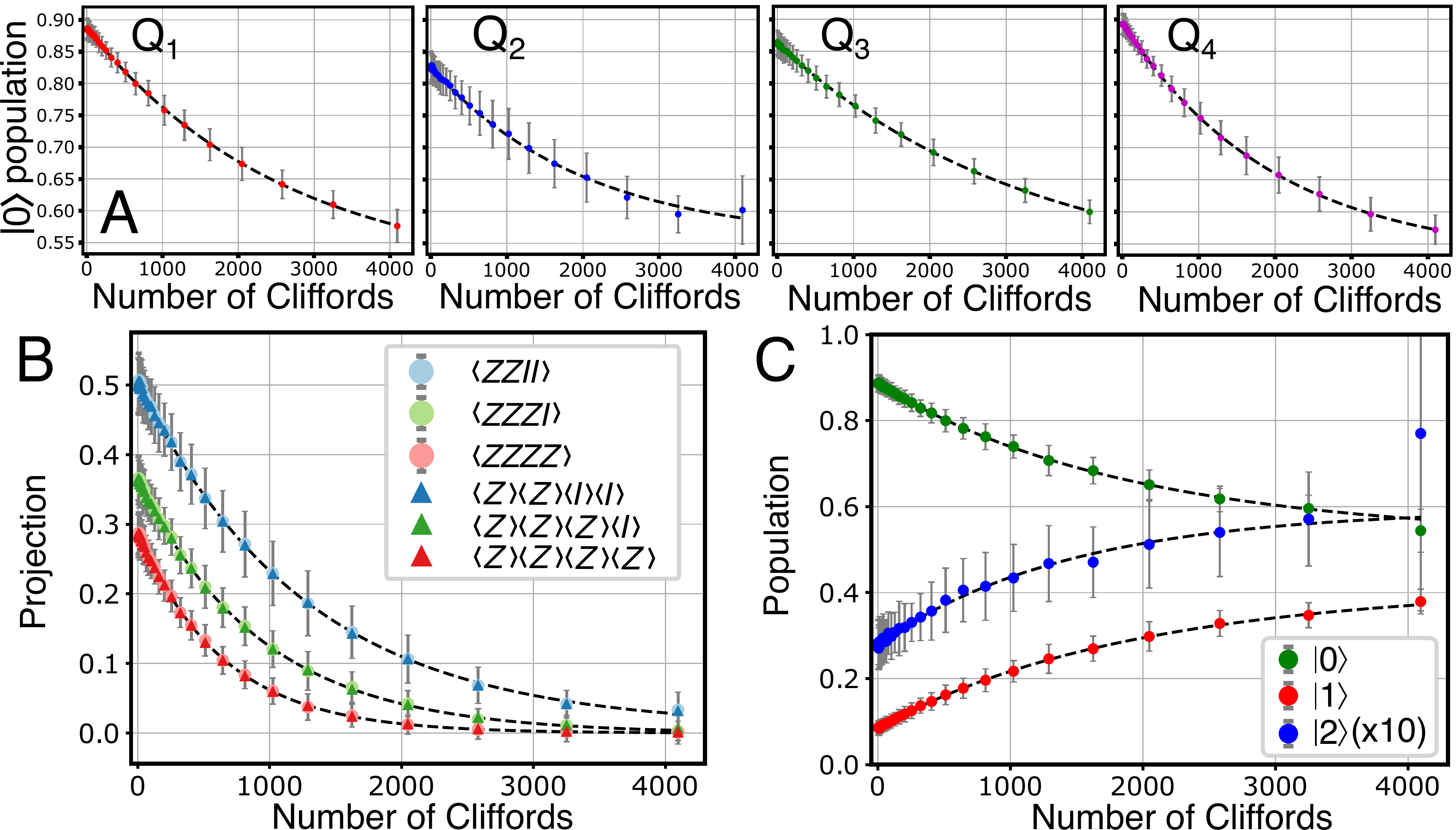}% Here is how to import EPS art
\caption{\label{fig:RB} \textbf{Randomized benchmarking.} (\textbf{A}) RB curves for simultaneous single-qubit RB on the four qubits. Points and error bars are the average and standard deviation of the results for the $k=80$ different Clifford sequences. (\textbf{B}) Pauli-$Z$ correlators $\langle ZZII\rangle$, $\langle ZZZI\rangle$, $\langle ZZZZ\rangle$ vs. number of Clifford gates for the single-shot simultaneous RB data. The fitted dashed curves provide the depolarizing parameters $\alpha_{1100}$, $\alpha_{1110}$, and $\alpha_{1111}$. The associated Pauli-$Z$ products vs. number of Clifford gates are also shown (triangles). The close similarity of the correlator and product curves is indicative of low crosstalk~\cite{mckay2020correlated,Gambetta2012}. (\textbf{C}) Leakage RB curve on $Q_3$. The final anomalous data point is excluded from the fit.}
\end{figure*}
Ramsey experiments were then performed on qubit $i$ ($i\neq j$) to measure the parasitic AC Stark-shift $\omega_{AC,i}$ as shown in Fig.~\ref{fig:crosstalk_characterization}(C). No AC Stark-shift $\omega_{AC,i}$ was detected for any combination of qubit $i$ and resonator $j$, with a frequency resolution of approximately \SI{1}{\kilo\hertz}, resulting in the approximate bound $\chi_{ij}/2\pi<\SI{20}{\hertz}$ using the dispersive relation $\omega_{AC,i}=2\chi_{ij} \bar{n}_j$~\cite{Gambetta2006}.
\section{Single qubit gate errors}
Single qubit randomized benchmarking (RB)~\cite{Chow2009,Gambetta2012} was performed on all four qubits both separately and simultaneously, using a combination of $\SI{24}{\nano\second}$ duration ($\SI{20}{\nano\second}$ Blackman envelope with $\SI{4}{\nano\second}$ buffer) physical $I,X_{\pi/2,\pi}$ gates with DRAG pulse shaping~\cite{Motzoi2009}, and virtual $Z$ gates~\cite{McKay2017}. Single-shot readout was performed for all the RB experiments (see Materials and Methods). Fig.~\ref{fig:RB}(A) shows the fitted RB curves for the simultaneous RB experiment. The RB protocol was run at $31$ Clifford sequence lengths and for $k=80$ different sequences of Clifford gates. 
\begin{figure}
\includegraphics[width=0.49\textwidth,height=0.49\textheight,keepaspectratio]{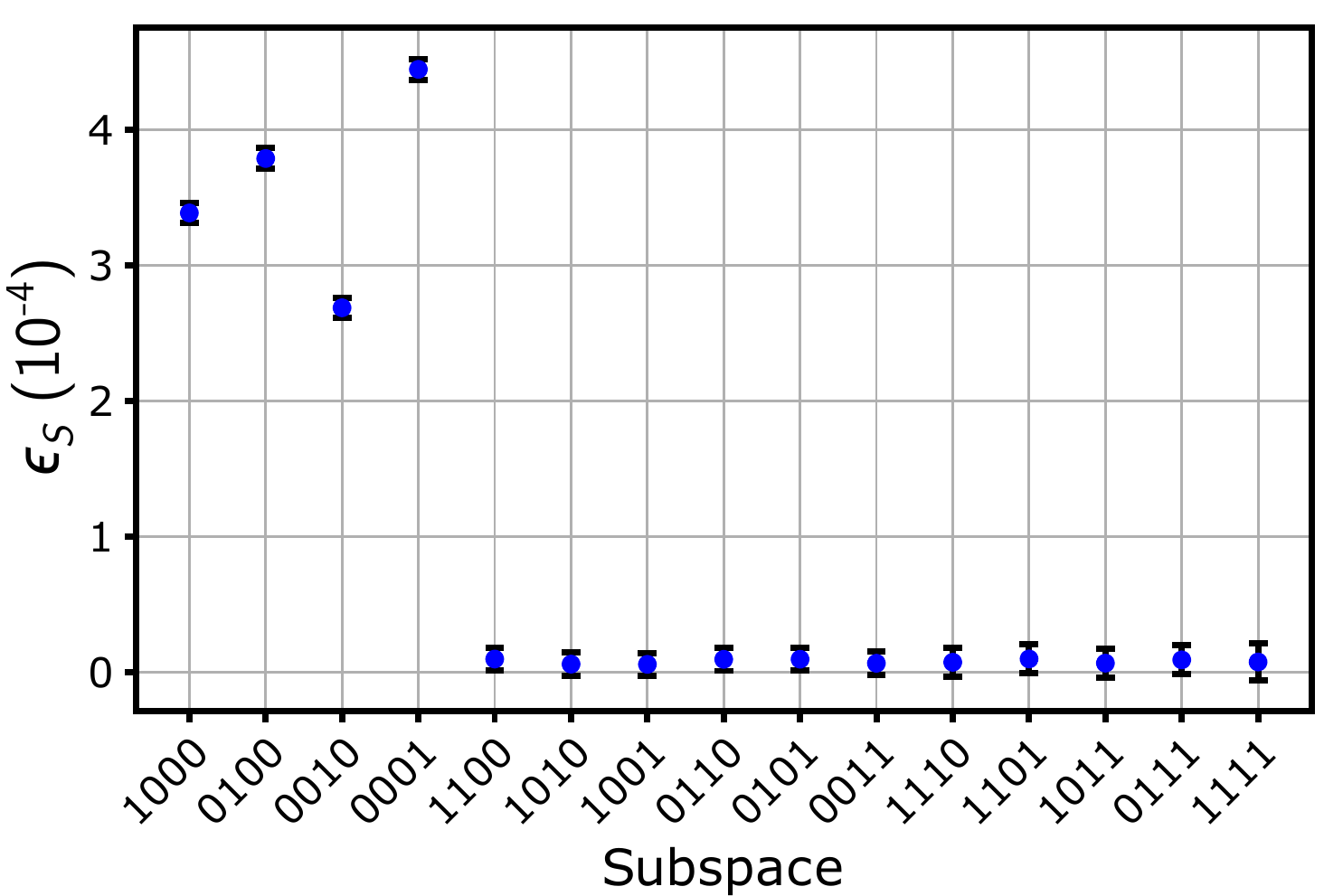}% Here is how to import EPS art
\caption{\label{fig:Factored_depol_error_per_Clifford} \textbf{Depolarizing fixed-weight parameters.} The four-qubit system depolarizing fixed-weight parameters $\epsilon_S$ in each subspace $S$ for $S\neq \varnothing$. The 15 subspaces are expressed as bitstrings where if bit $n$ (here indexed left to right) is 1 then qubit $n$ is in that subspace.}
\vspace{-1em}
\end{figure}
\begin{figure*}
\includegraphics[width=\textwidth,height=\textheight,keepaspectratio]{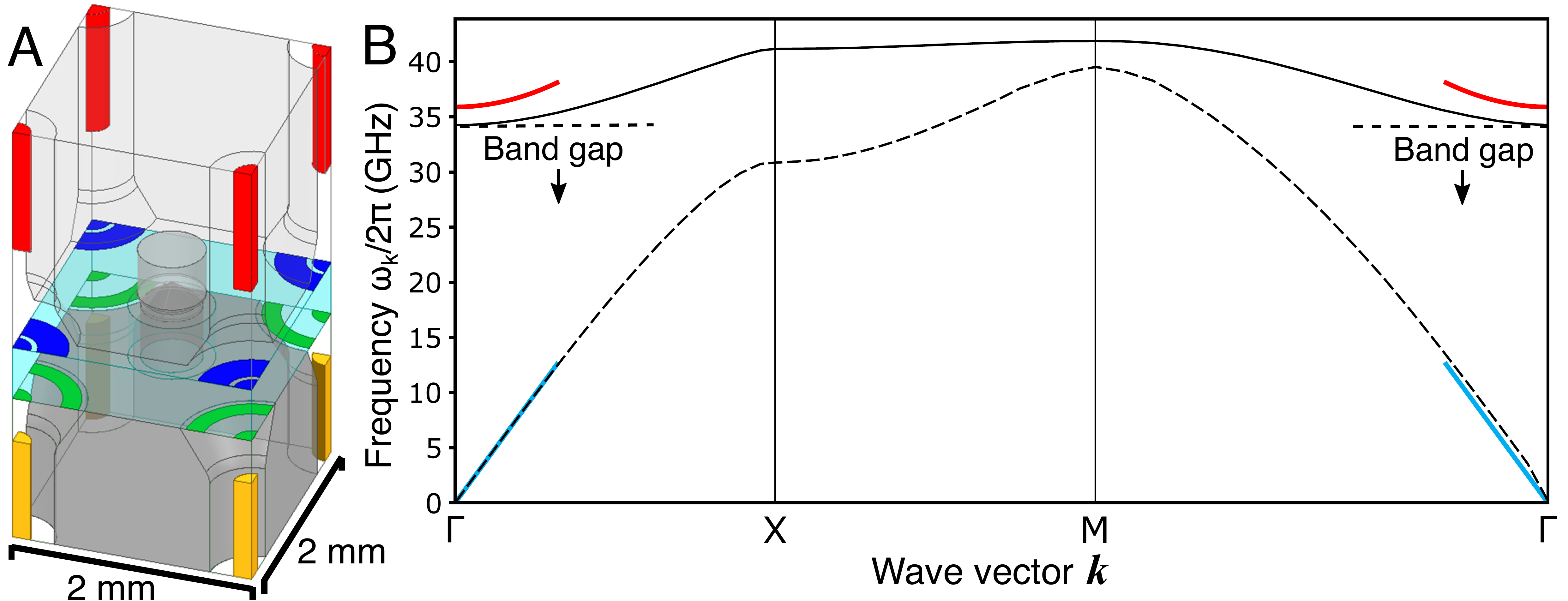}% Here is how to import EPS art
\caption{\label{fig:Band_structure} \textbf{Band structure simulation.} (\textbf{A}) HFSS model of a unit cell featuring a single addressable and measurable qubit ($4 \times 1/4$), and a single pillar that inductively shunts the enclosure. The unit cell has identical dimensions to the $\SI{2}{\milli\meter} \times \SI{2}{\milli\meter} $ central region of the device measured in this work. (\textbf{B}) Simulated lowest-band dispersion for the infinite enclosure formed by tiling the plane with the unit cell, with (solid) and without (dashed) the inductively shunting pillar and associated substrate aperture. The wave vector $\boldsymbol{k}$ traces between the symmetry points $\Gamma:(k_x=0,k_y=0)$, $X$:$(k_x=\pi/a,k_y=0)$, $M:(k_x=\pi/a,k_y=\pi/a)$. The colored curves show the predicted curvature around the $\Gamma$ point with (red) and without (blue) the inductively shunting pillar and associated substrate aperture, using no free fitting parameters (see Appendices).}
\end{figure*}
Each of the $31\times80$ experiments were repeated $5000$ times to build statistics. The resulting error-per-physical-gates (EPG) are presented in Table 2.
\\
Correlated RB (CorrRB)~\cite{mckay2020correlated} was performed using the simultaneous RB experiment data. Fig.~\ref{fig:RB}(B) shows a selection of the $2^4-1$ Pauli-$Z$ correlators vs. Clifford sequence length $m$. These were fit to standard RB curves $A\alpha_S^m+B$, where $S \subseteq B_4,S\neq \varnothing ,B_4=\{\{0\},\{1\},\{2\},\{3\}\}$ as defined in Ref.~\cite{mckay2020correlated}. From the fitted depolarizing parameters $\alpha_S$, the depolarizing fixed-weight parameters $\epsilon_S$ and the crosstalk metric $\tilde{\eta}$ were calculated~\cite{mckay2020correlated}. The $\epsilon_S$ parameters can be interpreted as the probability of a depolarizing error occurring in subspace $S$ per Clifford gate, and $\tilde{\eta}$ is a scalar quantity that expresses the distance of the measured four-qubit error channel from the nearest product of single-qubit error channels~\cite{mckay2020correlated}. The calculated $\epsilon_S$ values are shown in Fig.~\ref{fig:Factored_depol_error_per_Clifford} and we \linebreak find $\tilde{\eta} \approx 1 \times10^{-4}$ (see Appendices).
\\
Leakage RB (LRB)~\cite{Wood2018} was performed separately on $Q_3$, which had the highest readout fidelity (see Materials and Methods). The resultant LRB curve is shown in Fig.~\ref{fig:RB}(C), with a leakage-per-physical-gate (LPG) of $3.49(7)\times10^{-5}$; and an EPG of $2(1)\times10^{-4}$ found using a four fit parameter model, and $2.33(7)\times10^{-4}$ found using a more robust three fit parameter model that assumes $\text{EPG}\gg\text{LPG}$~\cite{Wood2018}.
\section{Band structure simulations}
Fig.~\ref{fig:Band_structure}(A) shows an HFSS model of a unit cell that has dimensions exactly matching the ideal dimensions of the central $\SI{2}{\milli\meter} \times \SI{2}{\milli\meter}$ region of the device measured in this work. Fig.~\ref{fig:Band_structure}(B) shows the simulated lowest-band dispersion of the infinite structure formed by tiling the plane with this unit cell, for both cases that inductively shunting pillar and associated substrate aperture are included and not included in the unit cell. In the case that the pillar is excluded, the band spans from $\SI{0}{\giga\hertz}$ to $\SI{39.5}{\giga\hertz}$ and undesired frequency collisions between qubits and this band are guaranteed. In contrast, when the pillar is present the band has a cut-off frequency of $\omega_c/2\pi=\SI{34.3}{\giga\hertz}$, with a band gap below extending to $\SI{0}{\giga\hertz}$. The simulated curvature around the cutoff frequency, defined $\omega_k=\omega_c+Ak^2$ where $k=\sqrt{k_x^2+k_y^2}$, is $A/2\pi=\SI{4.5}{\giga\hertz  \milli\meter \squared}$. A plasma metamaterial model~\cite{Pendry1996, Pendry1998} can be applied to the infinite structure~\cite{Spring2020, Murray2016} to predict a cutoff frequency at $\omega_c/2\pi=\SI{35.9}{\giga\hertz}$ and a band curvature of $A/2\pi=\SI{8.8}{\giga\hertz \milli \meter \squared}$, where these predictions neglect dissipation. This same metamaterial model can be used to predict that the spatial dependence for cavity mediated transverse coupling between equal frequency qubits takes the form~\cite{Spring2020}:
\begin{equation}
J_{cav,ij}=aK_0 (d_{ij}/\delta_p)\text{.}
\label{eq:J_spatial_dependence}
\end{equation}
Here, $a$ is a spatially independent term, $K_0$ is the modified Bessel function of the second kind, $d_{ij}$ is the spatial separation between qubits $i$ and $j$, and $\delta_p=1/\sqrt{\mu_0 \epsilon_0 \epsilon_r (\omega_c^2-\omega_q^2) }$ is the plasma skin depth. Using the simulated value $\omega_c/2\pi=\SI{34.3}{\giga\hertz}$ results in a predicted plasma skin depth of $\delta_p\approx\SI{0.7}{\milli\meter}$ for the unit cell considered here, assuming $\omega_q\ll\omega_c$. The spatial dependence tends to $J_{cav,ij}\propto e^{-d_{ij}/\delta_p }/\sqrt{d_{ij}/\delta_p }$ for $d_{ij}\gg\delta_p$. This equates to $J_{cav,ij}$ decreasing by approximately $\SI{25}{\decibel}$ for each $\SI{2}{\milli\meter}$ increase in qubit separation. Cavity mediated qubit (resonator)-control line couplings $\varepsilon_{cav,ij}^{q(r)}$ and qubit-resonator transverse couplings $g_{cav,ij}$ ($i \neq j$) are likewise predicted to have the same spatial dependence.
\\
The band structure was mapped out using HFSS, with details on the simulation model as well as the analytical cutoff frequency, band curvature, and plasma skin depth predictions provided in the Appendices.
\section{Discussion}
The architecture presented in this paper uses an inductively shunted cavity enclosure that tightly surrounds the circuit, combined with 3D integrated out-of-plane control wiring and ‘reverse-side’ readout resonators. The results demonstrate that this design is compatible with transmon relaxation times $T_1$ at least in the range of $\SI{150}{\micro\second}$ to $\SI{200}{\micro\second}$. The observed variation in $T_1$ is consistent with measured variation in transmon qubits on hour-long time scales and is suggestive of coupling to two-level system (TLS) defects~\cite{Mueller2019} as the dominant relaxation mechanism~\cite{Burnett2019,Klimov2018,Mueller2015}. The marked residual excited state population of the qubits suggests quasi-particle induced relaxation may also be significant~\cite{Serniak2018,Catelani2011}, indicating a potential need for improved infra-red filtering of signals to the device~\cite{Serniak2019,Barends2011}. The radiatively limited $T_1$ time of qubits in this device is predicted to be $\sim\SI{5}{\milli \second}$ using HFSS simulations~\cite{nigg2012black, houck2008controlling} (see Appendices). The architecture is also demonstrated to be compatible with pure echoed dephasing times $T_{\phi,e}$ of at least $\SI{180}{\micro\second}$. The average measured $T_{\phi,e}$ values bound the residual photon number and temperature of the four readout resonators to $\bar{n}_{th}\leq 6.5\times10^{-3}$ and $T_{r}\leq \SI{80}{\milli\kelvin}$ \cite{Wang2019}. A possible topic for further work is to clarify the effect of mechanical vibrations in the out-of-plane wiring on qubit dephasing.
\\
The results further establish that the architecture exhibits low crosstalk and can transmit short $\SI{20}{\nano\second}$ control pulses that execute single-qubit gates of high fidelity $F \approx 99.98\%$. The average gate fidelity was the same within error for the separate and simultaneous RB, implying that crosstalk errors were inconsequential at the measured fidelity. The small value of the crosstalk metric $\tilde{\eta} \approx 1 \times 10^{-4}$, and the small values of $\epsilon_S$ for weight $|S|>1$ show that depolarizing errors with weight $|S|>1$ were highly suppressed. The average error per gate was approximately $50\%$ coherence limited, and the leakage per gate as characterized on $Q_3$ was found to be less than $20\%$ of the error per gate. These values might be improved in future by more detailed pulse shaping and phase error correction~\cite{McKay2017}.
\\
A shortcoming of the presented device is that it exhibited small external resonator decay rates $\kappa_{ext,i}$ and dispersive shifts $\chi_{ii}$ that were non-optimal for qubit readout~\cite{Heinsoo2018, Walter2017}. The small $\kappa_{ext,i}$ values may be attributed to slight movement of the control line inner conductors due to material contraction during cooling to cryogenic temperatures. The small $\chi_{ii}$ values were due to the large qubit-resonator detunings and the choice of qubit and resonator electrode dimensions. We anticipate that future devices can achieve improved readout parameters.
\section{Conclusions}
In this work, average trasmon qubit coherence times of $T_1=\SI{149(38)}{\micro\second}, T_{\phi,e}=\SI{189(34)}{\micro\second}$ and simultaneous single-qubit gate fidelities of $F= 99.982(4)\%$ have been measured in a four-qubit demonstration of a 3D integrated superconducting circuit architecture. It has been shown that, prior to the inclusion of qubit coupling circuitry, residual crosstalk is highly suppressed. It is anticipated that a unit cell inside the device can be tiled to form larger devices that feature lattices of qubits. Band structure simulations predict that such devices will possess a cutoff frequency to cavity modes that is well above qubit frequencies, in agreement with a metamaterial model that further predicts cavity mediated crosstalk between qubits in these lattices will decay exponentially with spatial separation. A potential near-term application for this architecture is the study of correlated errors generated by high energy radiation \cite{wilen2021correlated, martinis2021saving}, where correlations could be probed in lattices of qubits with high coherence and exponentially suppressed crosstalk.
\section{Materials and methods}
The device enclosure was CNC machined from 6061 aluminum with $\pm\SI{10}{\micro\meter}$ machining tolerance on features. The out-of-plane wiring was made from silver plated copper (SPC) UT47 coaxial cable. The outer jackets and dielectrics were stripped back to expose the inner conductors (see Fig.~\ref{fig:device_assembly}). The circuit was fabricated on a double-side polished high-resistivity intrinsic silicon wafer using a double-sided waferscale process. Following a hydrofluoric acid dip, aluminum was deposited onto both sides of the wafer by evaporation. The qubit and resonator electrodes were defined by a wet etching process, and the qubit Josephson junctions were formed using the Dolan Bridge double-angle shadow evaporation technique~\cite{Dolan1988}. After circuit fabrication, the wafer was diced into square dies with side lengths of $4975\pm \SI{15}{\micro\meter}$ using a Disco DAD3430 dicing saw, and a $\SI{650}{\micro\meter}$ diameter aperture was then CNC drilled in the center of selected dies using a Loxham Precision $\si{\micro}6$ micro-machining system. An approximately $\SI{0.5}{\micro\meter}$ thick layer of S1805 photoresist was used as a protective layer during these CNC processes.
\\
Qubit readout was performed using a standard heterodyne detection technique~\cite{wallraff2005approaching}. It was possible to perform simultaneous single-shot readout on the four qubits using $\SI{5}{\micro\second}$ measurement pulses, with assignment fidelities: $F=\{97.8,97.7,98.5,98.4\}\%$, where $F=1-p(e|g)-p(g|e)$~\cite{Walter2017}.
\\
To extract the frequency of Ramsey fringes in Ramsey experiments, an interpolation method was applied to improve the frequency resolution found from the Fourier transform of the time traces~\cite{Gasior2004}. Details are provided in the Appendices.
\\
Experiments were all carried out during a single cooldown inside an Oxford Instruments Triton 500 dilution refrigerator, with a base stage temperature of $\sim \SI{20}{\milli\kelvin}$. A diagram of the dilution refrigerator setup is included in the Appendices.
\begin{acknowledgments}
This work has received funding from the United Kingdom Engineering and Physical Sciences Research Council under Grants No. EP/M013243/1, EP/N015118/1, EP/T001062/1, EP/M013294/1 and from Oxford Quantum Circuits Limited. T.T. acknowledges support from the Masason Foundation and the Nakajima Foundation. S.F. acknowledges support from the Swiss Study Foundation and the Bakala Foundation. B.V. acknowledges support from an EU Marie Sklodowska-Curie fellowship. We would like to thank Loxham Precision for their support with milling parameters and tooling.
\end{acknowledgments}
\appendix
\renewcommand\thefigure{A\arabic{figure}}
\renewcommand\thetable{A\Roman{table}}
\setcounter{figure}{0}
\setcounter{table}{0}
\section{Dilution refrigerator setup}
\begin{figure}
\includegraphics[width=0.49\textwidth,height=0.49\textheight,keepaspectratio]{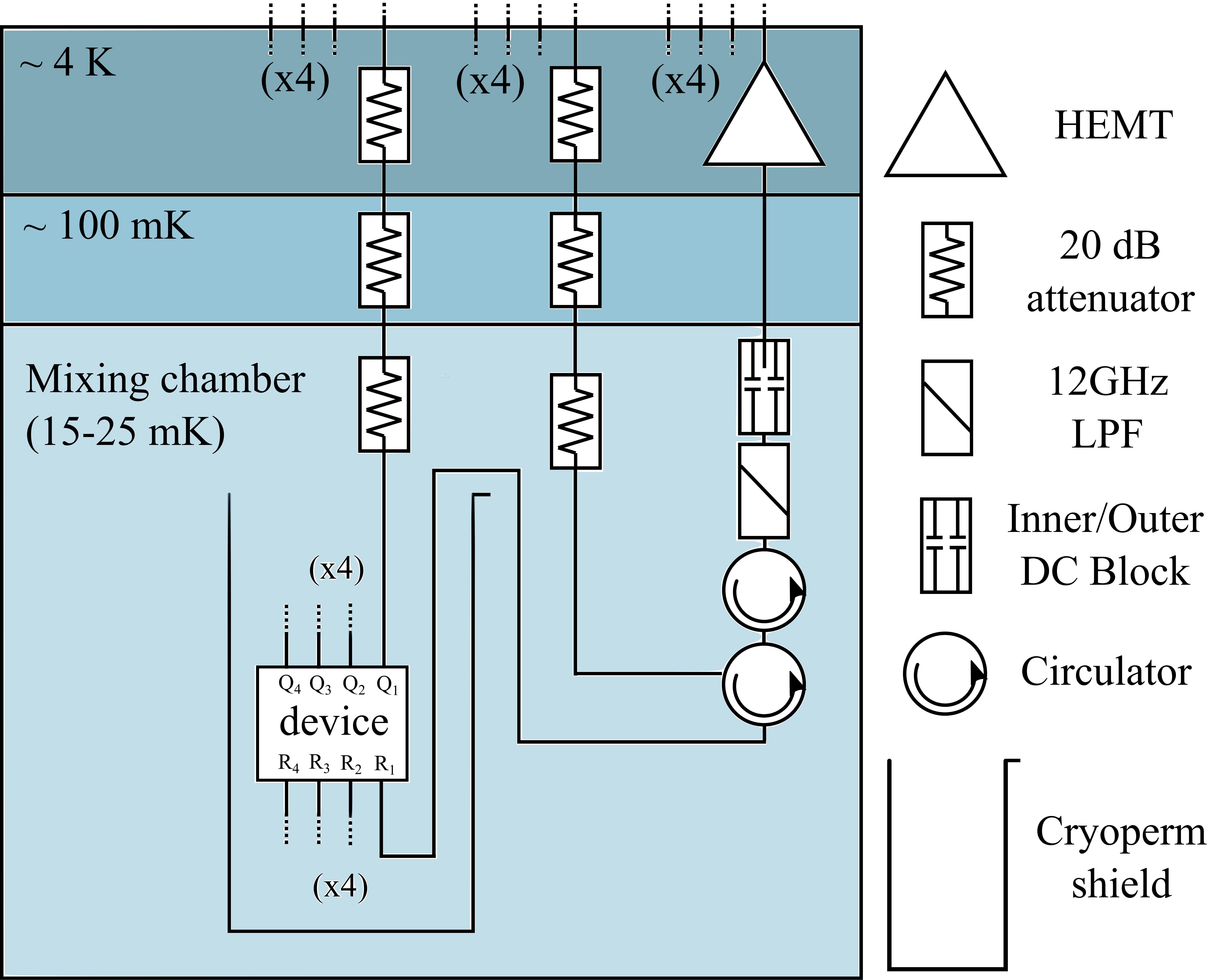}% Here is how to import EPS art
\caption{\label{fig:fridge_setup} \textbf{Dilution refrigerator setup.} Control wiring diagram of the dilution refrigerator used in the experiments.}
\end{figure}

Fig.~\ref{fig:fridge_setup} shows the experimental setup of the Oxford Instruments Triton 500 dilution refrigerator.
\begin{figure*}
\includegraphics[width=0.85\textwidth,height=0.85\textheight,keepaspectratio]{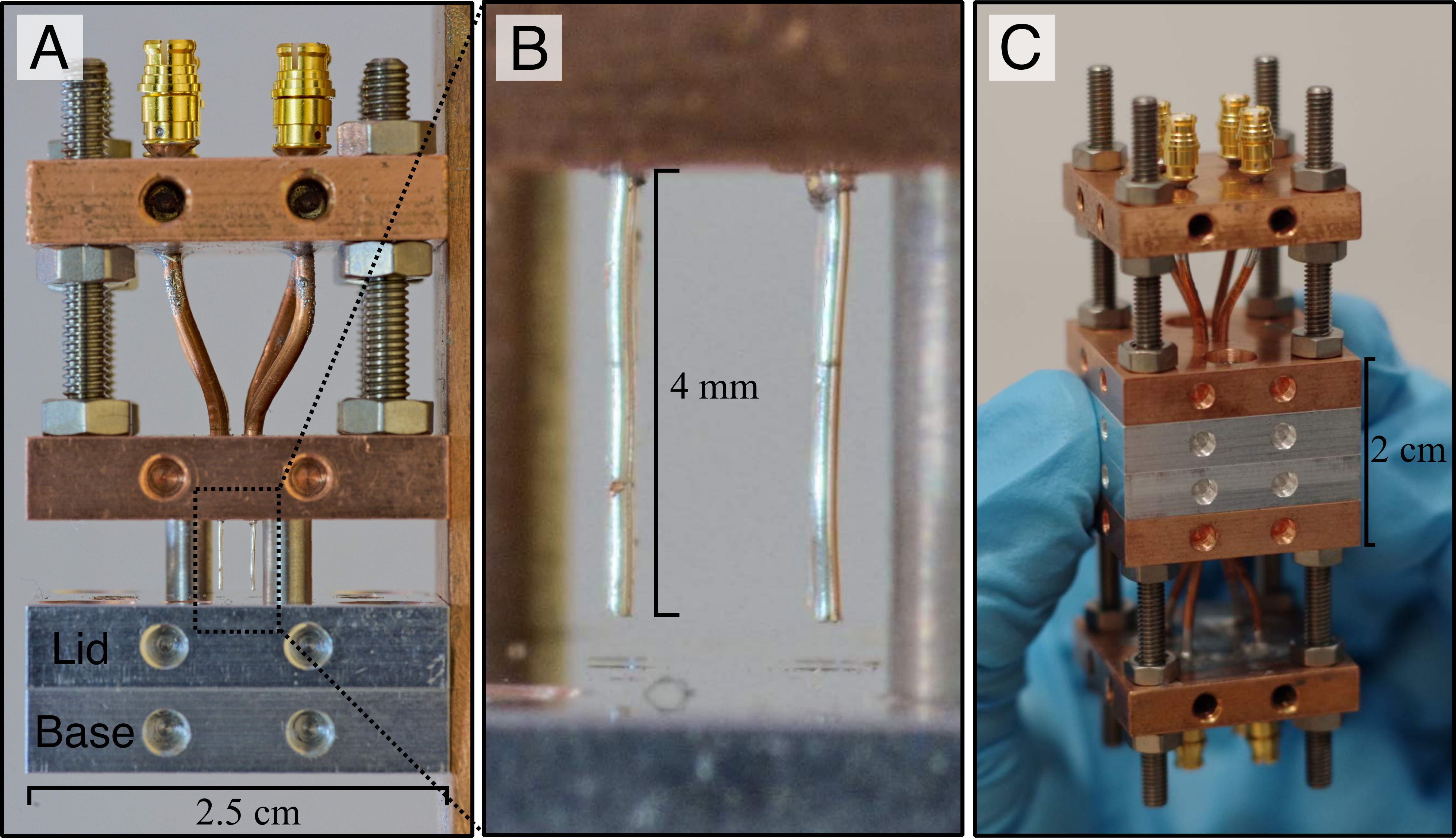}% Here is how to import EPS art
\caption{\label{fig:device_assembly} \textbf{Optical images of device enclosure and control wiring.} (\textbf{A}) The sealed base and lid of the enclosure, with a qubit control wiring piece suspended above. The UT47 coaxial cables are visible, terminated with SMP connectors to interface with the fridge wiring. (\textbf{B}) Zoomed in view of the extended inner conductors of the UT47 coaxial cables used for qubit control. (\textbf{C}) Fully assembled device, with the control wiring pieces for both qubits and resonators attached.}
\end{figure*}
\section{Device enclosure and control wiring}
Fig.~\ref{fig:device_assembly} shows optical images of the device. In Fig.~\ref{fig:device_assembly}(A), the sealed base and lid parts are shown with the wiring piece for qubit control partially inserted such that the protruding inner conductors are visible. The wiring piece is made from oxygen-free copper to improve device thermalization. The coaxial cables that deliver control signals to qubits and resonators are non-magnetic silver-plated copper (SPC) UT47 coaxial cables with an outer diameter of \SI{1.2}{\milli\meter}. At the fridge-facing end, the coaxial cables are terminated with SMP connectors, which can be arranged into a lattice with a higher packing density than SMA connectors. At the circuit-facing end, the exposed inner conductors of these coaxial cables, having a diameter of $\SI{0.29}{\milli\meter}$, extend out a distance $d_2\approx\SI{4}{\milli\meter}$ from the copper plate, where $d_2$ is defined in Fig.~\ref{fig:Device_schematics}(A) of the main text. Each wiring piece is aligned to the lid and base parts by steel dowel pins. Fig.~\ref{fig:device_assembly}(B) shows an enlarged view of the extended inner conductors used for qubit control. The through holes in the lid that these inner conductors slot into are also visible, having a diameter of $\SI{0.7}{\milli\meter}$. A lateral misalignment of the inner conductors greater than $\SI{0.2}{\milli\meter}$ will result in the inner conductors shorting to the device enclosure, causing the out-of-plane wiring to fail. Fig.~\ref{fig:device_assembly}(C) shows the fully assembled device, with the control wiring pieces for qubit control and resonator control both attached. The fasteners are made from titanium.
\section{Dephasing characterization measurements}
\begin{figure}
\includegraphics[width=0.5\textwidth,height=0.5\textheight,keepaspectratio]{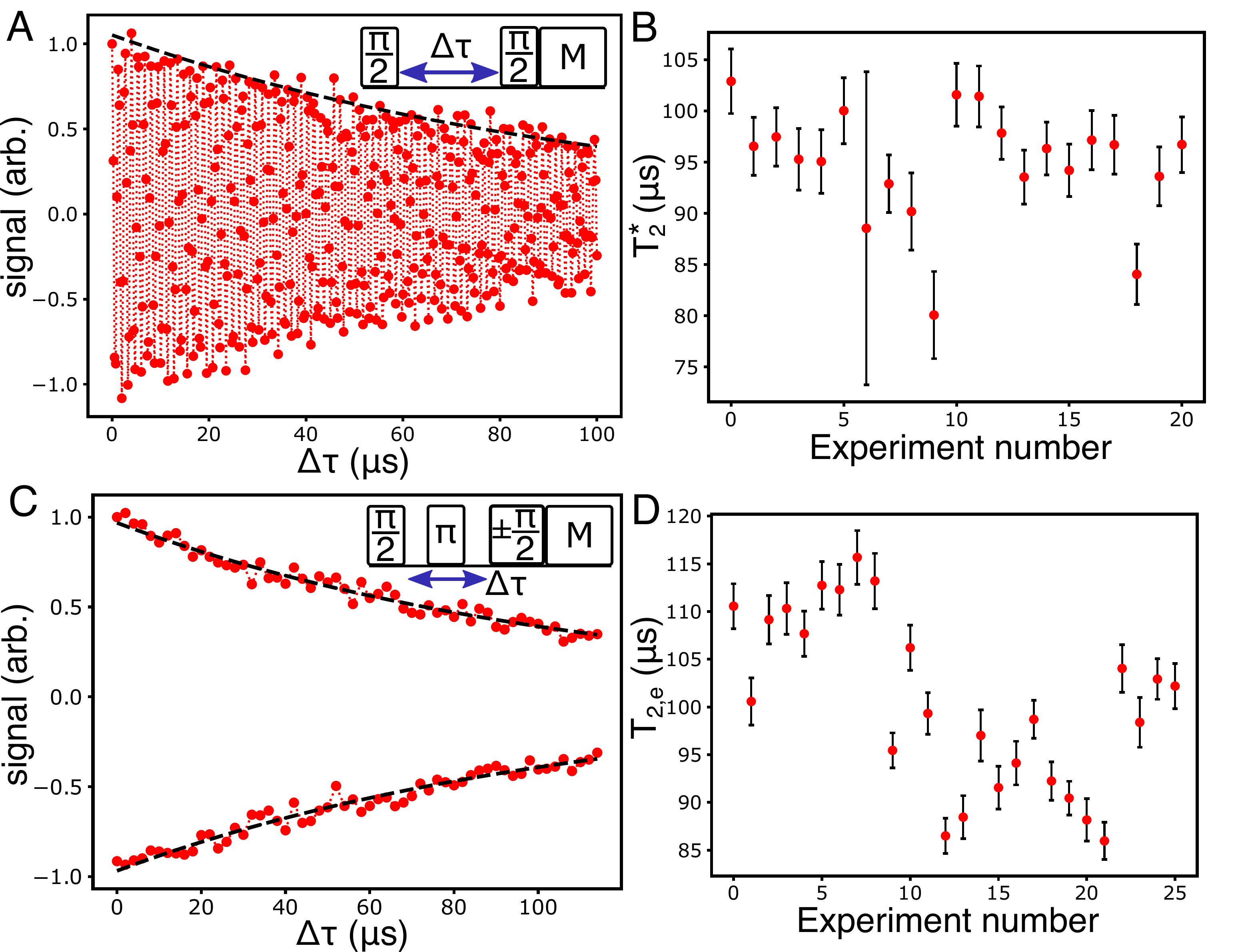}
\caption{\label{fig:T2_measurement_data} $\boldsymbol{T_2}$ \textbf{measurements on} $\boldsymbol{Q_1}$\textbf{.} (\textbf{A}) Example Ramsey time trace measurement with fitted exponential decay. The pulse sequence is shown in the inset. (\textbf{B}) Consecutive measured $T_2^*$ values found from repeated Ramsey time trace measurements. (\textbf{C}) Example Hahn echo time trace measurement with fitted exponential decay. The pulse sequence is shown in the inset. (\textbf{D}) Consecutive measured $T_{2,e}$ values found from repeated Hahn echo time trace measurements.}
\end{figure}

Example $T_2$ Ramsey and $T_2$ Hahn echo experiments that were performed on $Q_1$ are shown in Fig.~\ref{fig:T2_measurement_data}. The pulse sequences for these experiments are shown in the insets. $T_2$ Ramsey experiments were repeated 21 times on each qubit taking approximately 2 hours per qubit. $T_2$ Hahn echo experiments were repeated 26 times on each qubit taking approximately 2 hours per qubit. Each Hahn echo experiment was repeated with an opposite phase on the final $\pi/2$ qubit control pulse, resulting in two exponential decay curves as shown in Fig.~\ref{fig:T2_measurement_data}(C). These two curves were fit to $S=A\text{exp}(-\Delta\tau/T_{2,e})+B$ and $S=-A\text{exp}(-\Delta\tau/T_{2,e} )+B$ respectively, which improved the robustness of the extracted fit parameters A, B and $T_{2,e}$.
\section{Qubit control line selectivity}
\begin{figure*}
\includegraphics[width=0.9\textwidth,height=0.9\textheight,keepaspectratio]{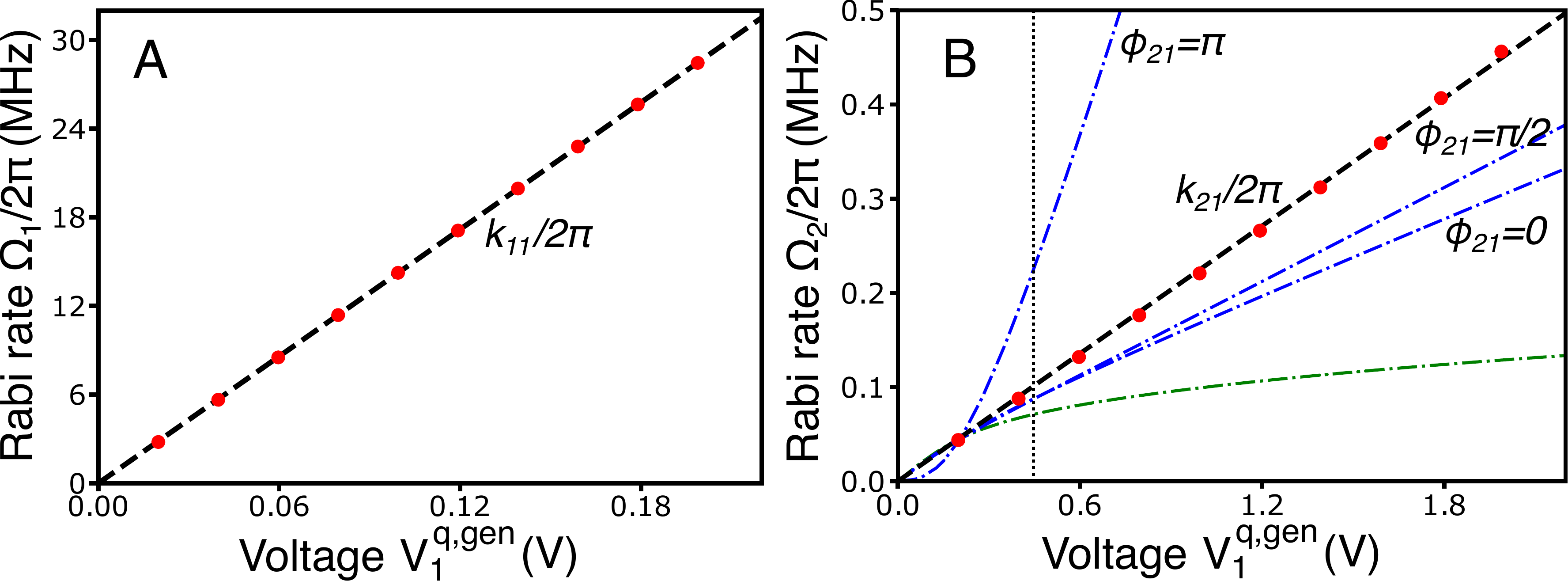}
\caption{\label{fig:qubit_control_line_selectivity_measurement} \textbf{Qubit crosstalk measurement data.} (\textbf{A}) Experimentally measured Rabi rate in $Q_1$ over a range of drive voltages through qubit control line 1, with $\omega_{d}=\omega_{q,1}$. The dashed black line shows a linear fit to the data. (\textbf{B}) Experimentally measured Rabi rate in $Q_2$ over a range of drive voltages through qubit control line 1, with $\omega_{d}=\omega_{q,2}$. The dashed black line shows a linear fit to the data. The thin dotted black line indicates the drive voltage where $|\varepsilon_{11}^q V_1^q|=|\Delta_{21}^q|$. The blue and green dot-dashed lines shown some predicted responses outside the $|J_{21}/\Delta_{21}^q |\ll |\varepsilon_{21}^q/\varepsilon_{11}^q|$ regime, described in the text.}
\end{figure*}

Fig.~\ref{fig:qubit_control_line_selectivity_measurement} shows the measured Rabi oscillation rates $\Omega_i$ in $Q_1$ and $Q_2$ when continuously driving these qubits on resonance through qubit control line 1 over a range of drive voltages $V_1^{q,gen}$. The fitted linear responses $\Omega_1=k_{11} V_1^{q,gen}$ and $\Omega_2=k_{21} V_1^{q,gen}$ are also shown. 
\\
The voltage $V_i^q$ at qubit control line $i$ satisfies $V_i^q=\lambda_i^q V_i^{q,gen}$ where $\lambda_i^q$ takes into account the attenuation between the room temperature generator and the device. 
It is assumed that $\lambda_i^q$ and $\varepsilon_{ij}^q$ are independent of the drive frequency over the range of qubit frequencies.
\\
In the main text, it is stated that from the good fit of the Rabi rate $\Omega_i$ to the function $\Omega_i=k_{ij} V_j^{q,gen}$ in the strong drive regime $|\varepsilon_{jj}^q V_j^q| \gg|\Delta_{ij}^q |$, it follows that $|J_{ij}/\Delta_{ij}^q |\ll|\varepsilon_{ij}^q/\varepsilon_{jj}^q|$, where $\Delta_{ij}^q=\omega_{q,i}-\omega_{q,j}$. This is now discussed. An effective expression for the total drive Hamiltonian on qubit $i$ from qubit control line $j$ is given by
\begin{equation}
\hat{H}_{D,ij} = V_j^q[\varepsilon_{ij}^q \hat{X}+\varepsilon_{ij}^J(\hat{X}\cos\phi_{ij} +\hat{Y}\sin\phi_{ij})]\text{,}
\end{equation}
where in addition to the control line crosstalk $\varepsilon_{ij}^q$, $\varepsilon_{ij}^J$ describes a coupling between qubit $i$ and qubit control line $j$ mediated through the transverse coupling $J_{ij}$ between qubits $i$ and $j$ (and which is in general dependent on the state of qubit $j$). The phase $\phi_{ij}$ accounts for any phase difference between $\varepsilon_{ij}^J$ and $\varepsilon_{ij}^q$. A drive voltage $V_j^q$ at frequency $\omega_d=\omega_{q,i}$ will result in Rabi oscillations in qubit $i$ with a rate $\Omega_i$ given by
\begin{equation}
    \Omega_i=V_j^q [(\varepsilon_{ij}^q)^2+(\varepsilon_{ij}^J)^2+2\varepsilon_{ij}^q \varepsilon_{ij}^J  \cos \phi_{ij} ]^{1/2}\text{.}
    \label{eq:general_driven_rabi_rate}
\end{equation}
In the low drive strength regime $|\varepsilon_{jj}^q V_j^q|\ll |\Delta_{ij}^q|, \varepsilon_{ij}^J$ is approximately independent of $V_j^q$, and for qubit $j$ in the ground state and $\omega_d=\omega_{q,i}$ is given by~\cite{Tripathi2019}
\begin{eqnarray}
    \varepsilon_{ij}^J = \varepsilon_{jj}^q\frac{J_{ij}}{\Delta_{ij}^q}[1-2(\frac{\varepsilon_{jj}^qV_j^q}{\Delta_{ij}^q})^2+4\frac{(\varepsilon_{jj}^qV_j^q)^2}{\Delta_{ij}^q(2\Delta_{ij}^q-\alpha_j)}+ \nonumber
    \\ \mathcal{O}((\varepsilon_{jj}^qV_j^q)^3)]\text{.} \;\;\;\;\;
\end{eqnarray}
In the high drive strength regime $|\varepsilon_{jj}^q V_j^q|>|\Delta_{ij}^q |$, $\varepsilon_{ij}^J$ is dependent on the drive voltage $V_j^q$, and can be numerically solved for using the semianalytical method introduced in Ref.~\cite{Tripathi2019}. The dashed green curve in Fig.~\ref{fig:qubit_control_line_selectivity_measurement}(B) shows the predicted relationship between $V_1^{q,gen}$ and $\Omega_2$ using this semianalytical method to determine $\varepsilon_{21}^J$, assuming that $|\varepsilon_{21}^J|\gg |\varepsilon_{21}^q|$ (i.e. $|J_{21}/\Delta_{21}^q |\gg|\varepsilon_{21}^q/\varepsilon_{11}^q |$) and that $Q_1$ is in the ground state. The semianalytical model Hamiltonian is truncated to the first 10 levels, and $J_{21}$ is treated as a fit parameter, which is chosen such that the curve intersects the first data point. The obvious poor fit of this curve to the data is clear evidence that $|J_{21}/\Delta_{21}^q|\gg|\varepsilon_{21}^q/\varepsilon_{11}^q |$ is \textit{not} satisfied in the measured device. The blue curves show the predicted response for $|J_{21}/\Delta_{21}^q |=|\varepsilon_{21}^q/\varepsilon_{11}^q |$, for phase values $\phi_{ij}=0,\pi/2,\pi$. Again, $J_{21}$ is chosen such that the curves intersect the first data point. In this case, the non-linear contribution of $\varepsilon_{21}^J$ still results in curves that differ significantly from the measured data. Since all of the measured Rabi rates show an excellent fit to the linear function $\Omega_i=k_{ij} V_j^{q,gen}$, we infer that $|\varepsilon_{ij}^q|\gg|\varepsilon_{ij}^J|$ and hence that $|J_{ij}/\Delta_{ij}^q |\ll|\varepsilon_{ij}^q/\varepsilon_{jj}^q |$, consistent with the simulation predictions in Table~\ref{tab:crosstalk_bounds}. In this case, eq.~\ref{eq:general_driven_rabi_rate} reduces to $\Omega_i=V_j^q \varepsilon_{ij}^q$, resulting in the following expression for $k_{ij}$
\begin{equation}
k_{ij}=\lambda_j^q \varepsilon_{ij}^q
\end{equation}
which leads to the relation in the main text $\varphi_{ij}^q=(k_{ij}/k_{jj} )^2$, where we restate for the reader that $\varphi_{ij}^q \stackrel{\text{def}}{=} (\varepsilon_{ij}^q/\varepsilon_{jj}^q)^2$.
\section{Resonator control line selectivity}
\begin{figure*}
\includegraphics[width=0.9\textwidth,height=0.9\textheight,keepaspectratio]{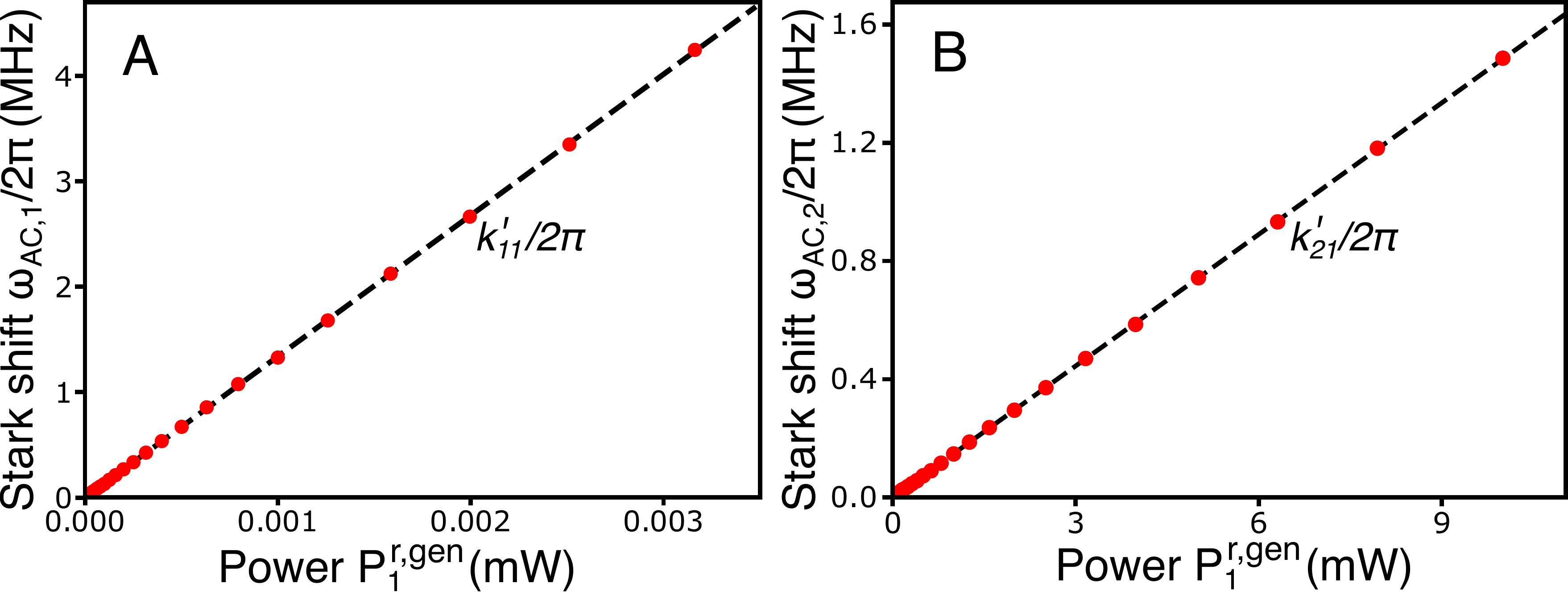}
\caption{\label{fig:resonator_control_line_selectivity_measurement} \textbf{Resonator crosstalk measurement data.} (\textbf{A}) Experimentally measured AC Stark shift in $Q_1$ over a range of drive powers through resonator control line 1, with $\omega_{d,1}=\omega_{r,1}+\Delta_d$ and $ \Delta_d/2\pi=\SI{5}{\mega\hertz}$. The dashed black line shows a linear fit to the data. (\textbf{B}) Experimentally measured AC Stark shift in $Q_2$ over a range of drive powers through resonator control line 1, with $\omega_{d,1}=\omega_{r,2}+\Delta_d$ and $\Delta_d/2\pi=\SI{5}{\mega\hertz}$. The dashed black line shows a linear fit to the data.}
\end{figure*}

Fig.~\ref{fig:resonator_control_line_selectivity_measurement} shows the measured AC-Stark shifts in $Q_1$ and $Q_2$ when continuously driving resonators $R_1$ and $R_2$ through resonator control line 1 at frequency $\omega_{d,1}=\omega_{r,1}+\Delta_d$ with $\Delta_d/2\pi=\SI{5}{\mega\hertz}$, over a range of drive powers $P_1^{r,gen}$. The fitted linear responses $\omega_{AC,1}=k_{11}' P_1^{r,gen}$ and $\omega_{AC,2}=k_{21}' P_1^{r,gen}$ are also shown. 
\\
The voltage $V_i^r$ at resonator control line $i$ satisfies $V_i^r=\lambda_i^r V_i^{r,gen}$ where $\lambda_i^r$ takes into account the attenuation between the room temperature generator and the device, and where $V_i^{r,gen}=(Z_0 P_i^{r,gen})^{1/2}$, with $Z_0=\SI{50}{\ohm}$. 
It is assumed that $\lambda_i^r$ and $\varepsilon_{ij}^r$ are independent of the drive frequency over the range of resonator frequencies.
\\
The average photon number in resonator $i$ due to a continuous drive through resonator control line $j$ at frequency $\omega_{d,i}$ is given by \cite{Gambetta2006,boissonneault2010improved,Blais2020Circuit}
\begin{equation}
    \bar{n}_i = \frac{(\varepsilon_{ij}^rV_j^r)^2}{(\kappa_i/2)^2+(\omega_{d,i}-\tilde{\omega}_{r,i})^2}\text{,}
    \label{eq:photon_number_vs_drive_voltage}
\end{equation}
where $\tilde{\omega}_{r,i}$ is the effective frequency of resonator $i$ taking into account shifts due to (e.g.) the state of qubits. In the low photon number regime $\bar{n}_i\ll n_{crit,i}$, where it is restated that $n_{crit,i}= (\Delta_{ii}/2g_{ii})^2$, $\tilde{\omega}_{r,i}$ is to good approximation independent of the photon number $\bar{n}_i$. For a drive frequency $\omega_{d,i}=\omega_{r,i}+\Delta_d$, $\omega_{d,i}-\tilde{\omega}_{r,i}$ in eq.~\ref{eq:photon_number_vs_drive_voltage} becomes $\Delta_d-\tilde{\Delta}_{r,i}$ where $\tilde{\Delta}_{r,i}=\tilde{\omega_{r,i}}-\omega_{r,i}$. If qubit $i$ is in the first excited state, then $\tilde{\Delta}_{r,i}=2\chi_{ii}$. For $|\Delta_d|\gg|2\chi_{ii}|$, $\bar{n}_i$ becomes effectively independent of the state of qubit $i$. Note that $\bar{n}_i$ is to an excellent approximation also independent of the state of other qubits, since $|\chi_{ki}| \ll |\chi_{ii}|$ for $k\neq i$. In the case that additionally $(\Delta_d)^2\gg(\kappa_i/2)^2$, eq.~\ref{eq:photon_number_vs_drive_voltage} takes the simple form
\begin{equation}
       \bar{n}_i = \frac{(\varepsilon_{ij}^rV_j^r)^2}{(\Delta_d)^2}\text{.}
       \label{eq:photon_number_vs_drive_voltage_detuned}
\end{equation}
In the regime $\bar{n}_i \ll n_{crit,i}$, the induced AC Stark-shift in qubit $i$ due to this population in resonator $i$ is given by $\omega_{AC,i}=2\chi_{ii} \bar{n}_i$~\cite{Gambetta2006}. Substituting eq.~\ref{eq:photon_number_vs_drive_voltage_detuned} and $\omega_{AC,i}=k_{ij}' P_j^{r,gen}$ into this formula leads to the following expression for $k_{ij}'$
\begin{equation}
    k_{ij}'=\frac{2Z_0(\lambda_j^r)^2\chi_{ii}(\varepsilon_{ij}^r)^2}{(\Delta_d)^2}
\end{equation}
which leads to the relation in the main text $\varphi_{ij}^r=(\chi_{jj}/\chi_{ii})(k_{ij}'/k_{jj}')$, where we restate for the reader that $\varphi_{ij}^r\stackrel{\text{def}}{=}(\varepsilon_{ij}^r/\varepsilon_{jj}^r)^2$.
\section{Qubit-resonator crosstalk}
In the experiment to bound the parasitic dispersive shift $\chi_{ij}$, resonator $i$ was populated with photons using a drive applied on resonance with its ground state frequency $\omega_{r,i}$. The generator power $P_{crit,i}^{r,gen}$ required to populate resonator $i$ with $\bar{n}_{i}$ photons using a continuous drive applied at frequency $\omega_{r,i}$ was calibrated in the following manner. $T_2$ Ramsey experiments were carried out on qubit $i$ while applying a continuous drive through resonator control line $i$ at frequency $\omega_{r,i}$, over a range of drive powers $P_i^{r,gen}$. 
\begin{figure}
\includegraphics[width=0.49\textwidth,height=0.49\textheight,keepaspectratio]{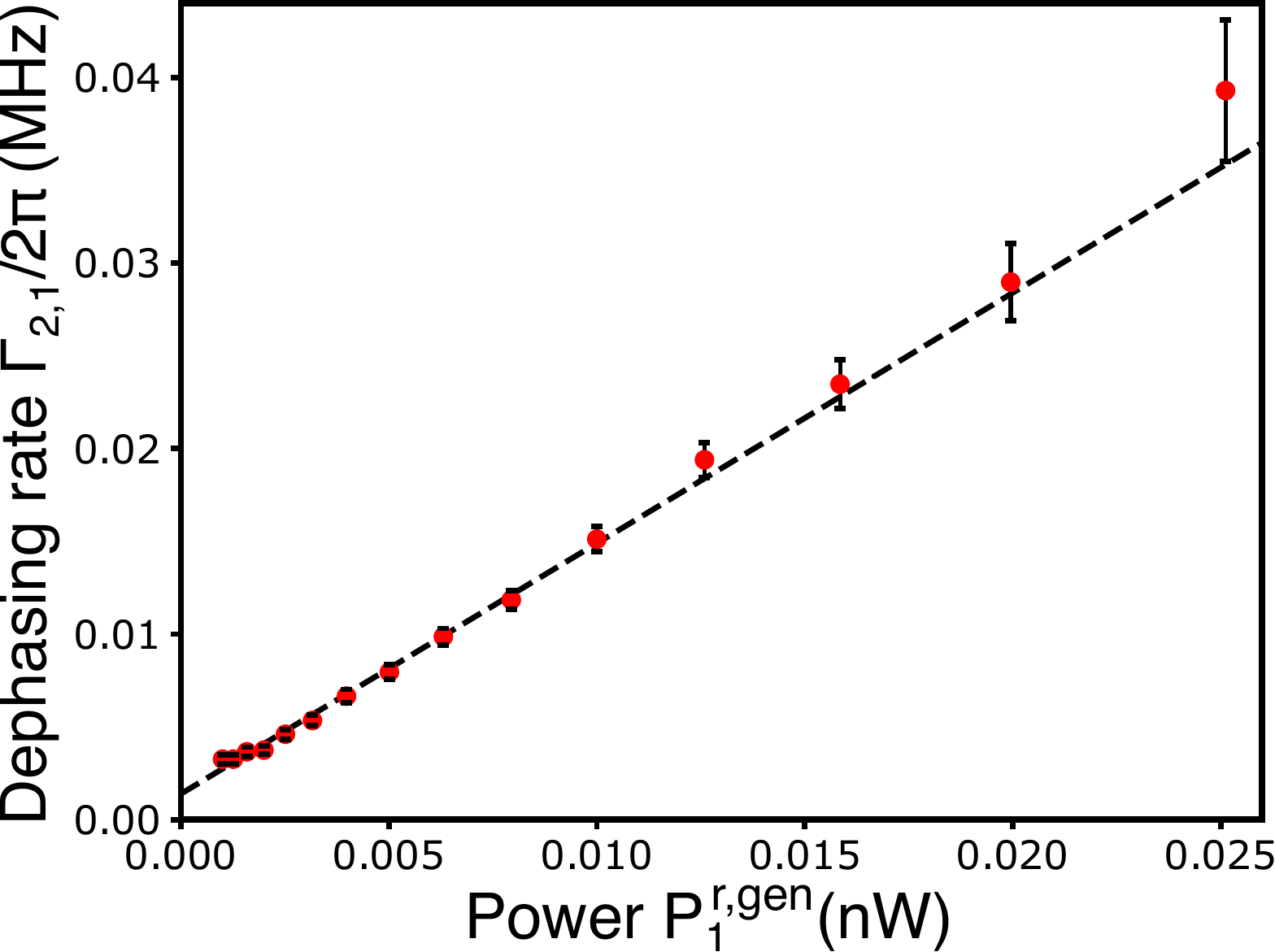}
\caption{\label{fig:dephasing_photon_number_calibration} \textbf{Dephasing data for photon number calibration.} Experimentally measured Ramsey dephasing rate $\Gamma_{2,1}=1/T_{2,1}^*$ in $Q_1$ over a range of drive powers through resonator control line 1, with $\omega_{d,1}=\omega_{r,1}$. The dashed black line shows the linear fit to the function $\Gamma_{2,1}=A_1+K_1 P_1^{r,gen}$.}
\end{figure}
The Ramsey decay rate $\Gamma_{2,i}=1/T_{2,i}^*$ was then fit to the formula $\Gamma_{2,i}=A_i+K_i P_i^{r,gen}$ as shown in Fig.~\ref{fig:dephasing_photon_number_calibration}. The photon number in resonator $i$ given qubit $i$ is in the ground state, $\bar{n}_{g,i}$, is (in the regime $\bar{n}_{g,i}\ll n_{crit,i}$) linearly related to the generator power through $\bar{n}_{g,i}=c_i P_i^{r,gen}$, where $c_i$ is given by
\begin{equation}
    c_i = \frac{\kappa_i [1+(4\chi_{ii}/\kappa_i)^2]K_i}{8(\chi_{ii})^2} \text{.}
    \label{eq:resonant_photon_number_drive_power_calibration}
\end{equation}
A derivation of eq.~\ref{eq:resonant_photon_number_drive_power_calibration} is provided here. The generalized measurement induced dephasing rate $\Gamma_{m,i}$ of qubit $i$ under a drive at frequency $\omega_{d,i}$ is given by~\cite{Gambetta2006, Blais2020Circuit}
\begin{equation}
    \Gamma_{m,i}=\frac{\kappa_i (\chi_{ii})^2 (\bar{n}_{e,i}+\bar{n}_{g,i})}{(\kappa_i/2 )^2+(\chi_{ii})^2+(\omega_{d,i}-[\omega_{r,i}+\chi_{ii}])^2} \text{,}
\end{equation}
where $\bar{n}_{e,i}$ is the average photon number in resonator $i$ given that qubit $i$ is in the excited state. For a drive at frequency $\omega_{d,i}=\omega_{r,i}$, the induced dephasing rate becomes
\begin{equation}
    \Gamma_{m,i}=\frac{8(\chi_{ii})^2\bar{n}_{g,i}}{\kappa_i[1 +(4\chi_{ii}/\kappa_i)^2]} \text{,}
    \label{eq:resonant_dephasing_vs_photon_number}
\end{equation}
where $\bar{n}_{e,i}=\bar{n}_{g,i}/[1+(4\chi_{ii}/\kappa_i)^2]$ has been used, which follows from eq.~\ref{eq:photon_number_vs_drive_voltage}. We stress that this expression and eq.~\ref{eq:resonant_dephasing_vs_photon_number} are only valid for a drive applied at frequency $\omega_{d,i}=\omega_{r,i}$. At times $t\gg1/\kappa_i$, the Ramsey dephasing rate of qubit $i$ under this dephasing drive is given by $\Gamma_{2,i}=\Gamma_i+\Gamma_{m,i}$~\cite{Gambetta2006, Blais2020Circuit}, where $\Gamma_i$ is the dephasing rate due to all other mechanisms independent of the drive. Making the association $\Gamma_{m,i}=K_i P_i^{r,gen}$ and substituting this along with $\bar{n}_{g,i}=c_i P_i^{r,gen}$ into eq.~\ref{eq:resonant_dephasing_vs_photon_number} then leads to eq.~\ref{eq:resonant_photon_number_drive_power_calibration}.
\\
Assuming for the moment that the expression $n_{g,i}=c_i P_i^{r,gen}$ remains valid outside the regime $\bar{n}_i\ll n_{crit,i}$, we calculate that the steady state photon numbers driven into each resonator in the qubit-resonator crosstalk experiment were: $\bar{n}_i=\{39(1),25(1),37(1),40(2)\}$, where $n_{crit,i}=\{258,257,251,256\}$. However, the simple linear expression $n_{g,i}=c_i P_i^{r,gen}$ is generally only valid in the limit $\bar{n}_i\ll n_{crit,i}$~\cite{Gambetta2006,boissonneault2010improved, Blais2020Circuit}. To determine a bound on parasitic dispersive shifts $\chi_{ij}$, we make the reasonable assumption that the linear relationship $\bar{n}_{g,i}=c_i P_i^{r,gen}$ remained valid in this device for $\bar{n}_i\leq n_{crit,i}/10$. Under this assumption, each resonator was driven with a photon number of at least $\bar{n}_i=n_{crit,i}/10$ (to within error) since the photon number is a monotonically increasing function with respect to the drive power. This leads to the bound $\chi_{ij}/2\pi<\SI{20}{\hertz}$ in the main text.
\section{Correlated RB depolarizing parameters and crosstalk metric}
\begin{figure}
\includegraphics[width=0.49\textwidth,height=0.49\textheight,keepaspectratio]{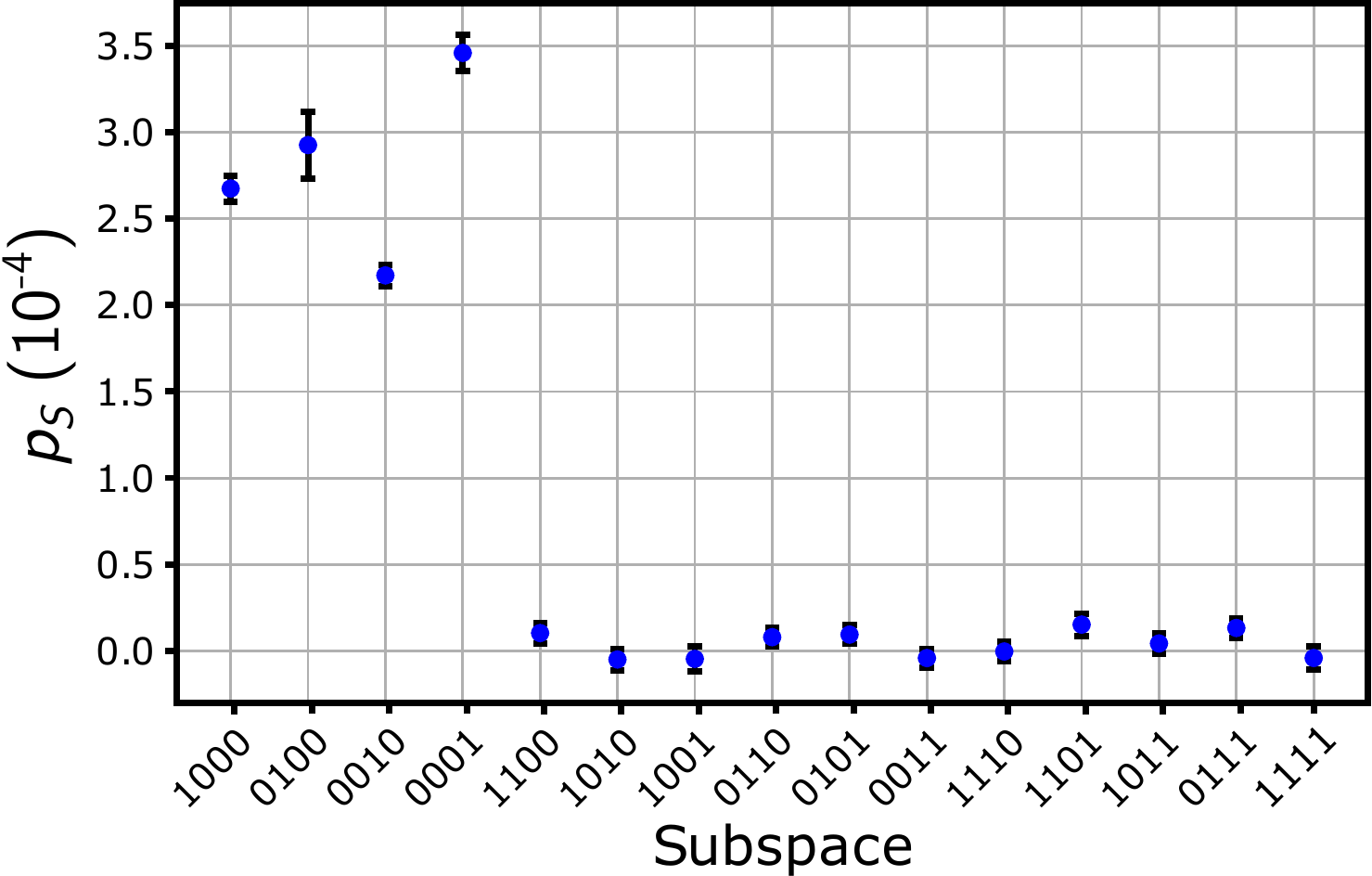}
\caption{\label{fig:Pauli_fixed-weight_parameters} \textbf{Pauli fixed-weight parameters.} The four-qubit system Pauli fixed-weight parameters $p_S$ in each subspace $S$ for $S\neq \varnothing$. The parameter $p_{S}$ for $S=0000$ (not shown on the graph) is equal to $0.99883(3)$.}
\end{figure}

The 15 $\alpha_S$ parameters determined by correlated RB are shown in Table.~\ref{tab:CorrRB_alpha_parameters}. The error bars were found by a bootstrapping technique. First, a collection of $100$ resampled RB data sets were generated from the the measured RB data set, by resampling with replacement from the $k=80$ different Clifford sequences. Correlated RB analysis was then performed on each of these resampled data sets, resulting in $100$ values for each $\alpha_S$ parameter, with the reported error being the standard deviation.
\\
The 15 $\alpha_S$ parameters were mapped directly onto 16 ($S=\varnothing$ is now included) Pauli fixed-weight parameters $p_S$ using a transformation without fitting parameters~\cite{mckay2020correlated}. The resulting $p_S$ values are shown in Fig.~\ref{fig:Pauli_fixed-weight_parameters}. These values should lie in the interval $[0,1]$ and satisfy $\sum_S p_S = 1$~\cite{mckay2020correlated}. The plotted error bars were estimated by mapping the collection of $100$ bootstrapped $\alpha_S$ parameter sets onto a collection of $100$ bootstrapped $p_S$ parameter sets, and taking the standard deviation of each $p_S$ parameter. The crosstalk metric was then calculated using the formula~\cite{mckay2020correlated}
\begin{equation}
    \tilde{\eta} = \sum_{S,|S|>1}|p_S| + \sum_{S,|S|\leq 1}|p_S-p_{S}'|,
\end{equation}
where $p_{S}'$ are Pauli fixed-weight parameters for an uncorrelated error channel such that $p_{S}'=0$ for $|S|>1$. These $p_{S}'$ are treated as fitting parameters, and are chosen to minimize $\tilde{\eta}$ while being contrained to lie in the interval $[0,1]$ and satisfy $\sum_{S,|S|\leq 1} p_{S}' = 1$. Since all the $p_S$ values should lie in the interval $ [0,1]$, the small negative $p_S$ values were set to zero when determining $\tilde{\eta}$. A collection of $100$ values of $\tilde{\eta}$ were calculated from the $100$ bootstrapped $p_S$ parameter sets and the standard deviation was used to estimate the error. The result was $\tilde{\eta} = 1.1(2)\times 10^{-4}$. We expect that the error determined by this method is an underestimate as it neglects errors introduced by the negative $p_S$ values. In the main text, we therefore report that $\tilde{\eta} \approx 1\times 10^{-4}$.
\begin{table}[b]
\caption{\label{tab:CorrRB_alpha_parameters} \textbf{Pauli-$Z$ depolarizing parameters.} The 15 fitted $\alpha_S$ depolarizing parameters determined from single-shot simultaneous RB, with standard deviation in brackets.}
\begin{ruledtabular}
\begin{tabular}{cc} 
Subspace $S$ & depolarizing parameter $\alpha_S$ \\
\hline
1000&0.99962(1) \\
0100& 0.99954(4)\\ 
0010& 0.99969(1)\\
0001& 0.99950(2)\\
1100& 0.99920(3)\\
1010& 0.99931(1)\\
1001& 0.99914(1)\\
0110& 0.99927(3)\\
0101& 0.99910(3)\\
0011& 0.99921(1)\\
1110& 0.99893(3)\\
1101& 0.99875(3)\\
1011& 0.99884(2)\\
0111& 0.99882(3)\\
1111& 0.99846(4)\\
\end{tabular}
\end{ruledtabular}
\end{table}
\begin{figure*}
\includegraphics[width=0.9\textwidth,height=0.9\textheight,keepaspectratio]{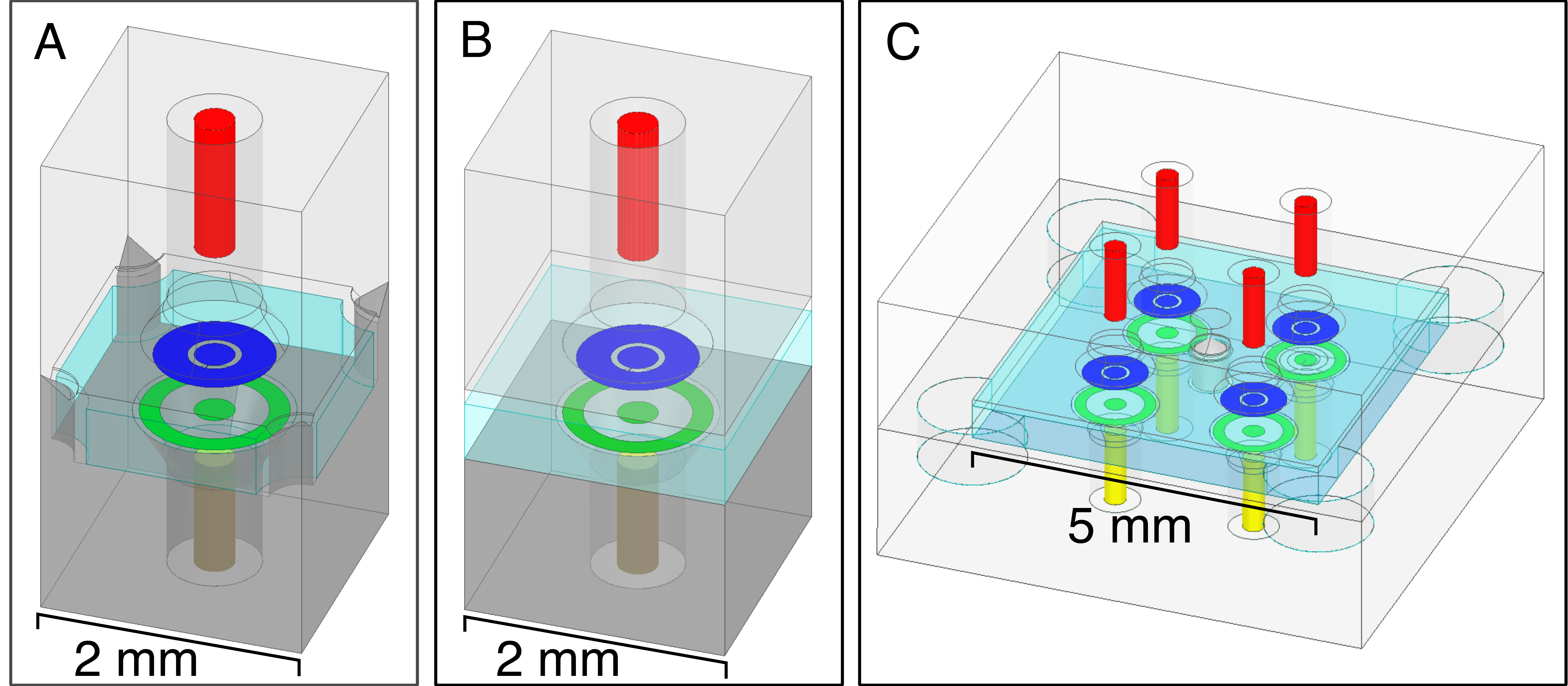}
\caption{\label{fig:HFSS_models} \textbf{HFSS models.} (\textbf{A}) The unit cell model with the inductive shunt. Perfect conductor is displayed in gray, silicon is displayed in light blue, and vacuum is displayed as transparent. The inner conductors of the qubit and resonator control lines (modeled as perfect conductor) are displayed in red and yellow. The qubit electrodes are displayed in dark blue and the resonator electrodes in green. (\textbf{B}) The unit cell model without the inductive shunt, following the same color convention. (\textbf{C}) The four-qubit device model used to predict radiative loss and parasitic transverse couplings, again following the same color convention.}
\end{figure*}
\section{Band structure simulations}
Fig.~\ref{fig:HFSS_models}(A)\&(B) show the HFSS models used to simulate the band structure in Fig. 7 of the main text. The enclosure was modeled as a perfectly conducting material. The relative permittivity of the silicon substrate was taken to be $11.45$~\cite{krupka2006measurements}, and the relative loss tangent was set to $0$, neglecting internal losses. The qubit and resonator electrodes were included in the simulation and modeled as perfectly conducting 2D sheets using the ‘perfect E’ boundary condition. The qubit and resonator control line inner conductors were modeled as perfectly conducting material. The model dimensions are such that the infinite structure formed by tiling this model is identical to that formed by tiling the model shown in Fig.~\ref{fig:Band_structure}(A) of the main text. In the model shown in  Fig.~\ref{fig:HFSS_models}(A) the control ports were terminated with $\SI{50}{\ohm}$ boundaries to simulate external losses out of the system. In the model shown in Fig.~\ref{fig:HFSS_models}(B) the control ports were terminated using the ‘perfect E’ boundary condition to avoid convergence issues encountered with the Eigenmode solver.
\\
Linked boundary condition (LBC) pairs were defined on the silicon and vacuum regions at the four faces of the models, using the so called ‘Master’ and ‘Slave’ boundary conditions. By changing the relative phase between Master and Slave pairs, the HFSS Eigenmode solver mapped out the band structure of the infinite structure formed by tiling the plane with the model.
\\
The analytical predictions for the plasma frequency and band-curvature around the $\Gamma$ point $(k_x=0,k_y=0)$, as well as the plasma skin depth, are discussed here. As a function of the radius $r$ and spacing $a$ of the pillar lattice, the plasma frequency $\omega_p$ was approximated using the following analytical formula~\cite{Belov2002, Krynkin2009}
\begin{equation}
    \omega_p = \frac{\omega_a}{\sqrt{\pi[\ln(a/r)-C]}}\text{,}
    \label{eq:analytic_plasma_frequency}
\end{equation}
where $\omega_a=\sqrt{2} \pi /(a\sqrt{\mu_0 \epsilon_0 \epsilon_r})$ and $C$ is a constant numerical factor approximately equal to $1.31$. The predicted dispersion around the $\Gamma$ point $(k_x=0,k_y=0)$ in the presence of the pillar lattice was found using the following expansion~\cite{Spring2020}
\begin{equation}
    \omega_k = \omega_p+\frac{1}{2\mu_0\epsilon_0\epsilon_r\omega_p}k^2+\mathcal{O}(k^4)\text{,}
\end{equation}
where $k=\sqrt{k_x^2+k_y^2}$. This results in $A=1/(2\mu_0 \epsilon_0 \epsilon_r \omega_p)$, where we restate for the reader that $A$ is defined by the relation $\omega_k=\omega_c+A k^2$. In the absence of the pillar lattice, the predicted dispersion around the $\Gamma$ point is given by $\omega_k=k/\sqrt{\mu_0 \epsilon_0 \epsilon_r}$, i.e. the dispersion of a plane wave propagating in the $x-y$ plane inside a medium with relative permittivity $\epsilon_r$. The relative permittivity $\epsilon_r$ for the enclosure was modified by the presence of the vacuum region above the silicon substrate. The thickness of the silicon substrate is $\SI{475}{\micro\meter}$ and the thickness of the vacuum region between the substrate and the enclosure lid is $\SI{125}{\micro\meter}$. Taking the cryogenic relative permittivity of silicon to be $11.45$, the effective relative permittivity was then calculated as~\cite{Spring2020}
\begin{equation}
    \epsilon_r = \frac{475+125}{475/11.45+125/1}\approx 3.6 \text{.}
\end{equation}
Inserting this into eq.~\ref{eq:analytic_plasma_frequency} with $a=\SI{2}{\milli\meter}$ and $r=\SI{0.25}{\milli\meter}$ results in the predictions $\omega_p/2\pi=\SI{35.9}{\giga\hertz}$ and $A/2\pi=\SI{8.8}{\giga \hertz \milli\meter \squared}$ that appear in the main text. Further inserting this value of $\epsilon_r$ into $\delta_p=1/\sqrt{\mu_0 \epsilon_0 \epsilon_r (\omega_p^2-\omega_q^2)}$ and taking $\omega_q\ll\omega_p$ results in the prediction $\delta_p\approx \SI{0.7}{\milli\meter}$ that also appears in the main text.
\\
We make the following observation regarding the effect of the vacuum region on cavity mediated crosstalk. In the regime $\omega_q\ll\omega_p$, the plasma skin depth is given by $\delta_p=a[\ln(a/r)-C]/\sqrt{2\pi}$, using eq.~\ref{eq:analytic_plasma_frequency}. This expression is independent of $\epsilon_r$. Thus, in the regime $\omega_q\ll\omega_p$, further reducing $\epsilon_r$ by increasing the thickness of the vacuum region is predicted to be an ineffective strategy for reducing the skin depth $\delta_p$ of cavity mediated crosstalk in this architecture.
\section{Radiative loss and transverse coupling simulations}
\begin{figure*}
\includegraphics[width=0.8\textwidth,height=0.8\textheight,keepaspectratio]{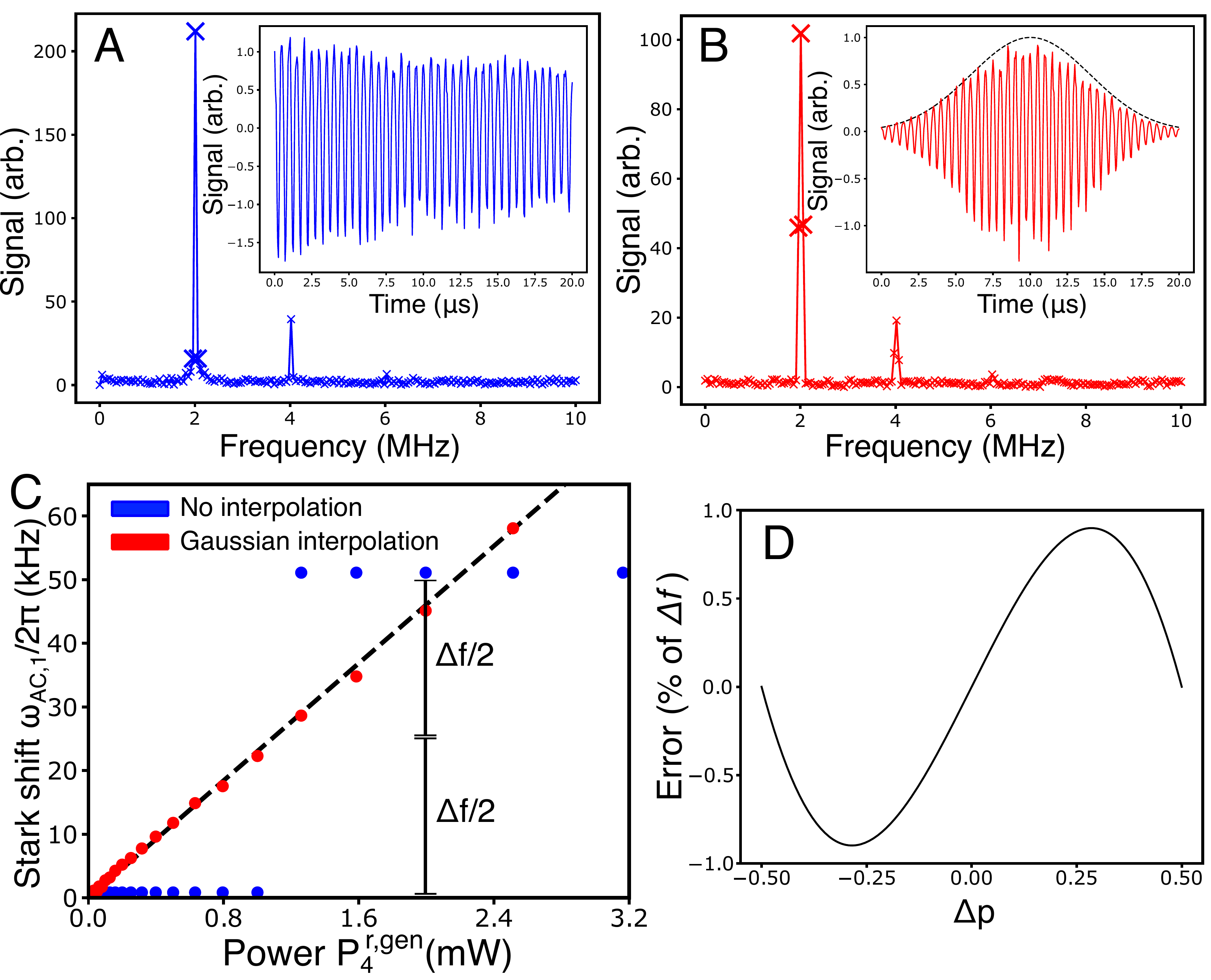}
\caption{\label{fig:fft_interpolation} \textbf{Ramsey experiment frequency resolution improvement of fft using Gaussian interpolation.} Ramsey Experiment on $Q_1$ while continuously driving $R_1$ through resonator control line 4 with generator power $P_4^{r,gen}$. (\textbf{A}) The fft of the $\SI{20}{\micro\second}$ long Ramsey time trace shown in the inset, for $P_4^{r,gen}=\SI{0.032}{\milli\watt}$. The crosses corresponding to the principal peak $p$ and the side peaks $p-1$ and $p+1$ are enlarged. (\textbf{B}) The fft of the same time trace data after windowing with a \SI{4}{\micro\second} standard deviation Gaussian function as shown in the inset. (\textbf{C}) The AC Stark shift $\omega_{AC,1}$ vs $P_4^{r,gen}$ found by: (blue) simple peak selection of the unwindowed fft, and (red) Gaussian interpolation of the windowed fft. (\textbf{D}) Predicted frequency error of the Gaussian interpolation method as a function of $\Delta p$ for $\sigma_{Guass}=T/5$, adapted from Ref.~\cite{Gasior2004}.} 
\end{figure*}
Fig.~\ref{fig:HFSS_models}(C) shows the HFSS model used to simulate the parasitic transverse couplings $J_{ij}$ and $g_{ij}$, and the radiatively limited relaxation time of qubits. The enclosure, silicon substrate, qubit and resonator electrodes, and control lines were all modeled identically to their counterparts in the band structure simulations. The control ports were terminated with $\SI{50}{\ohm}$ boundaries to simulate external losses out of the system.
\\
The Driven Terminal solver in HFSS was used to simulate the transverse couplings. In order to simulate $J_{ij}$ between qubits $i$ and $j$, all qubit electrode pairs were connected by lumped ports at the location of the Josephson junctions. The resonator electrode pairs were inductively disconnected to reduce simulation complexity. The transfer impedance $Z_{ij}$ between qubit ports $i$ and $j$ was simulated at a frequency of $\SI{4}{\giga\hertz}$. The transfer impedances were then inserted into a simple impedance formula~\cite{Solgun2019} to determine $J_{ij}$. To simulate $g_{ij}$ between qubit $i$ and resonator $j$, all resonator electrode pairs were directly connected by lumped ports (effects due to the spiral geometry of the resonator inductors were neglected), in addition to the lumped ports placed across qubit electrode pairs. The transfer impedance $Z_{ij}$ between qubit port $i$ and resonator port $j$ was then simulated at frequency values of $\SI{4}{\giga\hertz}$ and $\SI{8}{\giga\hertz}$ and the results were inserted into the same impedance formula to determine $g_{ij}$.
\\
In order to simulate the radiatively limited lifetime of qubits, one qubit in the model was connected by a lumped port at the location of its Josephson junction, and the electrodes of the associated resonator were connected by a spiral inductor [as visible in Fig.~\ref{fig:HFSS_models}(C)]. The remaining qubit and resonator electrode pairs were all inductively disconnected to reduce the simulation complexity. The impedance $Z$ at the qubit port was then simulated over a range of frequencies using the fast frequency sweep functionality, and the radiatively limited lifetime was found using the method in Ref.~\cite{nigg2012black}. The spiral inductor geometry was such that the simulated resonator frequency was $\omega_r/2\pi=\SI{8.2}{\giga\hertz}$, and the junction inductance was chosen such that $\omega_q/2\pi=\SI{4}{\giga\hertz}$.
\section{Ramsey interferometry frequency resolution improvement of fft using Gaussian interpolation}
To find the resonator control line selectivities $\varphi_{ij}^r$ and to bound the parasitic dispersive shifts $\chi_{ij}$, it was necessary to determine the dominant frequency in Ramsey experiment time trace data. Fig.~\ref{fig:fft_interpolation}(A) shows the fast Fourier transform (fft) of a $T=\SI{20}{\micro\second}$ long Ramsey time trace measurement on $Q_1$. The principal peak of the fft, index $p$, corresponds to the dominant oscillation frequency. The frequency separation of points in the fft, $\Delta f=1/T$, is in this case $\SI{50}{\kilo\hertz}$. The dominant frequency can be estimated as $\Delta f\times p$, with a resolution of $\Delta f/2$, which in this case is $\SI{25}{\kilo\hertz}$.
\\
Following Ref.~\cite{Gasior2004}, the frequency resolution was greatly improved by applying a Gaussian window with standard deviation $\sigma_{Guass}$ to the time trace data and then interpolating the ‘true’ dominant frequency value by using the fft signal amplitudes at the indices $p-1,p,p+1$, denoted $S_{p-1},S_p,S_{p+1}$. The predicted frequency is given by $f_{Gauss}=\Delta f(p+\Delta p)$ where $\Delta p$ satisfies~\cite{Gasior2004}
\begin{equation}
    \Delta p = \frac{\ln(S_{p+1}/S_{p-1})}{2\ln(S_p^2/[S_{p-1}S_{p+1}])} \text{.}
\end{equation}
Assuming that the time trace data is a sinusoid with frequency $f_0$, the frequency error of this Gaussian interpolation technique can also be found~\cite{Gasior2004}. The predicted value of the frequency error as a function of $\Delta p$ for $\sigma_{Guass}=T/5$ is plotted in Fig.~~\ref{fig:fft_interpolation}(D). The maximum error is $|f_0-f_{Guass} |=0.009\Delta f$. This corresponds to a frequency resolution of approximately $\SI{0.5}{\kilo\hertz}$ for $T=\SI{20}{\micro\second}$ as was used in the resonator control line selectivity experiments, and a maximum frequency resolution of approximately $\SI{0.1}{\kilo\hertz}$ for $T=\SI{90}{\micro\second}$ as was used in the parasitic dispersive shift bound experiments. Note that these predicted resolutions are smaller (i.e. better) than the true resolutions as they neglect noise and exponential decay in the Ramsey time trace data.
\\
The improvement in the frequency resolution afforded by this Gaussian windowing technique is shown in Fig.~\ref{fig:fft_interpolation}(B)\&(C). Fig.~\ref{fig:fft_interpolation}(B) shows the fft of the same time trace data as that in Fig.~\ref{fig:fft_interpolation}(A), where now the time trace data has been windowed with a Gaussian function with $\sigma_{Guass}=T/5$. Notice the far larger signal amplitudes $S_{p-1}$ and $S_{p+1}$ in this fft. Fig.~\ref{fig:fft_interpolation}(C) shows the AC Stark shift $\omega_{AC,1}$ while continuously driving $R_1$ through resonator control line 4 with generator power $P_4^{r,gen}$. Principal peak selection results in a frequency resolution of \SI{25}{\kilo\hertz}, whereas the Gaussian interpolation method achieves a frequency resolution of approximately $\SI{1}{\kilo\hertz}$. The linear fit to the interpolated frequencies provided the value of $k_{14}'$ which was used to determine the resonator control line selectivity $\varphi_{14}^r$.
% bib
\bibliography{complete_paper_references}% Produces the bibliography via BibTeX.
\end{document}